\documentclass[twocolumn]{aastex701}

\usepackage{bm}
\usepackage{amsmath}
\usepackage{graphicx}

\shorttitle{M60-UCD1 SMBH}
\shortauthors{Su, Li \& Hou}

\begin{document}

\title{Wind-fed Supermassive Black Hole Accretion in the Ultracompact Dwarf Galaxy M60-UCD1}

\author[0000-0002-6738-3259]{Zhao Su}
\affiliation{School of Astronomy and Space Science, Nanjing University, Nanjing 210046, China}
\affiliation{Key Laboratory of Modern Astronomy and Astrophysics (Nanjing University), Ministry of Education, Nanjing 210046, China}
\email[show]{suzhao@smail.nju.edu.cn}

\author[0000-0003-0355-6437]{Zhiyuan Li}
\affiliation{School of Astronomy and Space Science, Nanjing University, Nanjing 210046, China}
\affiliation{Key Laboratory of Modern Astronomy and Astrophysics (Nanjing University), Ministry of Education, Nanjing 210046, China}
\affiliation{Institute of Science and Technology for Deep Space Exploration, Suzhou Campus, Nanjing University, Suzhou 215163, China}
\email[show]{lizy@nju.edu.cn}

\author[0000-0001-9062-8309]{Meicun Hou}
\affiliation{Institute of Science and Technology for Deep Space Exploration, Suzhou Campus, Nanjing University, Suzhou 215163, China}
\email{houmc@nju.edu.cn}
\begin{abstract}
Ultracompact dwarf galaxies (UCDs) are thought to be remnants of stripped galactic nuclei, among which a handful are known to host a central supermassive black hole (SMBH). As in stripped nuclear star clusters, the SMBHs in UCDs may be fed by stellar winds from old stellar populations, in the absence of substantial gas reservoirs and galactic inflows. In this work, we investigate such a wind-fed accretion scenario for M60-UCD1, which harbors a confirmed $2\times10^7~M_\odot$ SMBH and exhibits X-ray emission suggestive of SMBH accretion signature. Using three-dimensional hydrodynamical simulations, we simulate the SMBH accreting stellar winds from approximately 1500 asymptotic giant branch stars, and explore the role of ram pressure from the ambient interstellar or intracluster medium. After 5 Myr, the majority of the stellar winds form a cold gas disk ($\sim1000~M_\odot$) within $\sim10~\rm pc$ as well as the SMBH’s gravitational sphere of influence. Within the inner $10^4~r_{\rm g}$, this disk transitions into a hot ($\sim10^7–10^9~\rm K$), geometrically thick corona that dominates the X-ray emission. The SMBH achieves an accretion rate of $\sim10^{-5}~M_\odot~\rm yr^{-1}$, yielding an X-ray luminosity of $\sim7\times10^{37}~\rm erg~s^{-1}$, well consistent with observations. Including ram pressure stripping reduces both the accretion rate and luminosity by about a factor of two. Our results suggest that the X-ray counterpart of M60-UCD1 originates from a weakly accreting SMBH fed by stellar winds, with broader insights into the feeding mechanisms of central massive black holes and the origins of X-ray sources in other UCDs.
\end{abstract}



\section{Introduction}\label{sec:intro}
Supermassive black holes (SMBHs) residing in galactic nuclei tend to have a low accretion rate throughout most of their lifetimes, observed as low luminosity active galactic nuclei \citep[LLAGNs;][]{2008ARA&A..46..475H}.
Although LLAGNs are suggested to have a capability for significantly influencing the evolution of their host galaxies, how the LLAGNs are fed remain poorly understand \citep{2017MNRAS.465.3291W, 2018MNRAS.479.4056W, 2019ApJ...885...16Y}.
Stellar winds from nuclear star clusters (NSCs; see recent review by \citealt{2020A&ARv..28....4N}), the dense and compact stellar systems prevalent in galaxy centers, could be one of the most immediate and direct supplies for the accretion onto these SMBHs \citep{2009ApJ...699..626H}.

Can these weakly accreting SMBHs be fed by stellar winds from NSCs?
The wind-fed accretion scenario has been explored for only a few nearby SMBHs but suggested to be promising (see the Introduction of \citealt{2025ApJ...988...68S} and references therein).
Notably, Sgr A*, the central SMBH of our own Galaxy, has been extensively studied in the context of the wind-fed scenario, benefiting from the well-constrained orbits of its surrounding massive stars \citep[e.g.,][]{2008MNRAS.383..458C,2018MNRAS.478.3544R,2020MNRAS.492.3272R,2020ApJ...896L...6R}.
In a previous work, we have demonstrated that wind-fed accretion can successfully explain the activity of M31* and the gas reservoir in its surrounding NSC, potentially representing a general case for other dormant SMBHs \citep{2025ApJ...988...68S}.
In addition to winds from NSCs, gas can be driven to the galactic center by various mechanisms on galactic scales, which complicates the feeding process and makes it challenging to quantify the role of NSC winds.
Therefore, galaxies lacking gas reservoirs and galactic-scale inflows provide``cleaner'' laboratories to examine and test the wind-fed scenario.

Ultracompact dwarf galaxies \citep[UCDs;][]{1999A&AS..134...75H,2000PASA...17..227D,2001ApJ...560..201P}, a peculiar class of small galaxies which to some can be viewed as an isolated super star cluster, may play such a role.
UCDs have predominantly old stellar populations and stellar masses ($\sim10^6-10^8~M_\odot$) comparable to faint dwarf elliptical galaxies, but exhibit extremely compact sizes of $\sim10-100~\rm pc$ \citep{2012MNRAS.425..325F, 2015ApJ...812...34L, 2020ApJS..250...17L}. 
It has been proposed that some UCDs, particularly those at the high-mass end, are remnants of dwarf galaxies tidally stripped by neighboring massive galaxies, leaving behind only their compact cores (e.g., NSCs) \citep[e.g.,][]{2001ApJ...552L.105B, 2003MNRAS.344..399B, 2003Natur.423..519D, 2013MNRAS.433.1997P, 2023Natur.623..296W, 2023MNRAS.526L.136P}.
Furthermore, UCDs and NSCs occupy overlapping regions in the stellar mass-effective radius plane \citep{2014MNRAS.443.1151N} and several UCDs have been found to host (super-)massive black holes \citep[e.g.,][]{2014Natur.513..398S,2018ApJ...858..102A}.
A key feature of UCDs in the context of wind-fed accretion is their gas-poor nature, as much of the gas in their progenitor galaxies is likely lost during the tidal stripping process, as evidenced by their predominantly old stellar populations and the observed absence of young stars \citep{2012MNRAS.425..325F,2015MNRAS.451.3615N,2016MNRAS.456..617J}.
In the absence of a significant gas reservoir, mass-loss from stellar populations in UCDs would be the primary gas supply.
Therefore, UCDs, serving as excellent analogues to ``naked" NSCs, provide a unique opportunity to apply and test the wind-fed accretion scenario.
M60-UCD1, the first UCD discovered to host an SMBH, is an ideal case for such a study, as it hosts the heaviest SMBH currently known in UCDs and is one of a few showing hints of SMBH accretion signature, namely an X-ray counterpart.
 
M60-UCD1, once considered as the brightest UCD and the densest galaxy, has a stellar mass of $1.2\times10^8~M_\odot$ and effective radius of 24 pc \citep{2013ApJ...775L...6S, 2014Natur.513..398S}.
It is located near the massive elliptical galaxy M60 (NGC\,4649) in the Virgo cluster, with a projected distance of about 7 kpc.
M60-UCD1 hosts an overmassive SMBH with a mass of $2.1\times10^{7}~\rm M_\odot$, which occupies 15\% of its total mass.
The X-ray counterpart of M60-UCD1 was first identified by \citet{2013ApJS..204...14L} using {\it Chandra} observations, in which it is named XID 144.
This source has an unabsorbed luminosity of $9.5\times10^{37}~\rm erg~s^{-1}$ in the 0.3--8 keV band and the observed X-ray flux displays significant variation over timescale of months to years, ranging from $5.6\times10^{37}~\rm erg~s^{-1}$ to $1.3\times10^{38}~\rm erg~s^{-1}$.
The X-ray spectrum is well fitted by an absorbed power-law spectrum with a photon index of 1.8.
Subsequent studies leveraging {\it Chandra} data have reported consistent results \citep{2016ApJ...819..162P, 2016ApJ...819..164H, 2018ApJ...858..102A, 2021MNRAS.506.4702F}.
The X-ray properties of the counterpart are consistent with those of a low-mass X-ray binary (LMXB), which is considered the most likely origin of the X-ray emission observed in general UCDs \citep{2016ApJ...819..162P, 2016ApJ...819..164H,2025ApJ...984..132F}.
The abundance of LMXBs in UCDs is notably higher than in field stellar populations but lower than in globular clusters. 
This has been linked to a top-heavy stellar initial mass function \citep{2012ApJ...747...72D} and to dynamical interactions within dense stellar systems \citep{2016ApJ...819..164H}. 
In addition to the LMXB hypothesis, accretion onto a putative central black hole presents an alternative origin, especially for the most massive UCDs, such as M60-UCD1.
\citet{2013ApJ...775L...6S} suggest that the X-ray source of M60-UCD1 can possibly be the accretion signature of an SMBH, which is justified by the relatively low possibility ($\sim25\%$) of an LMXB in M60-UCD1 at this luminosity.
If this is the case, the SMBH would has an inferred Eddington ratio ($L_{\rm bol}/L_{\rm Edd}$) of $\sim10^{-7}$, which places it in the regime of LLAGNs and is typical of galactic nuclei with old stellar populations \citep{2009ApJ...699..626H}.
Radio emission provides a useful tool to distinguish between the LMXB and SMBH scenarios, in that LMXBs are expected to emit much less in the radio band.
However, the deep radio observations with the VLA have yielded a non-detection for M60-UCD1, with an upper limit of $L_{\rm 5.8GHz}<1.14\times10^{34}~\rm erg~s^{-1}$ \citep{2018ApJ...858..102A}.
Moreover, this upper limit lies within the scatter of the fundamental plane of black hole activity for LLAGNs with $L_{\rm X}/L_{\rm Edd}<10^{-6}$ \citep{2017ApJ...836..104X}.
Consequently, it remains challenging to unambiguously associate the X-ray counterpart with SMBH accretion based on the current observations.

In this work, we apply the wind-fed accretion scenario for M60-UCD1 that is candidate for a stripped galaxy nucleus and among the most massive UCDs hosting the heaviest SMBHs.
The remainder of the paper is organized as follows. 
In Section \ref{sec:dynamics} and Section \ref{sec:simulation}, we describe the Schwarzschild dynamical modeling of M60-UCD1 and the setup of hydrodynamical simulations on the SMBH of M60-UCD1 fed by stellar winds, respectively.
In Section \ref{sec:results}, we present the simulation results and the impact of ram pressure from the interstellar and intracluster medium.
We discuss the feasibility of the wind-fed scenario for M60-UCD1 and the implications for active galactic nuclei in general UCDs in Section \ref{sec:discussion}.
We summarize our study in Section \ref{sec:conclusion}.

\section{Dynamical Modeling}\label{sec:dynamics}
To implement a self-consistent gravitational potential of M60-UCD1 in the simulation and model orbital motion of stars losing mass, we first construct an orbit-superposition Schwarzschild dynamical model following \citet{2014Natur.513..398S}.
The Schwarzschild method \citep{1979ApJ...232..236S} is commonly used to measure black hole dynamical mass and build triaxial dynamical models of galaxies.
This method will seek a dynamical solution with a large set of stellar orbits by simultaneously fitting the observed surface brightness and stellar kinematics of the galaxy.
We utilize the {\sc DYNAMITE} code\footnote{\url{https://github.com/dynamics-of-stellar-systems/dynamite}} \citep{2020ascl.soft11007J, 2022A&A...667A..51T}, the public version of an implementation of the Schwarzschild method by \citet{2008MNRAS.385..647V}.
Here, we briefly summarize the procedures and observation data used, which are detailed in \citet{2008MNRAS.385..647V} and \citet{2014Natur.513..398S}.

\begin{figure}[hbpt!]
	\centering
	\includegraphics[width=0.48\textwidth]{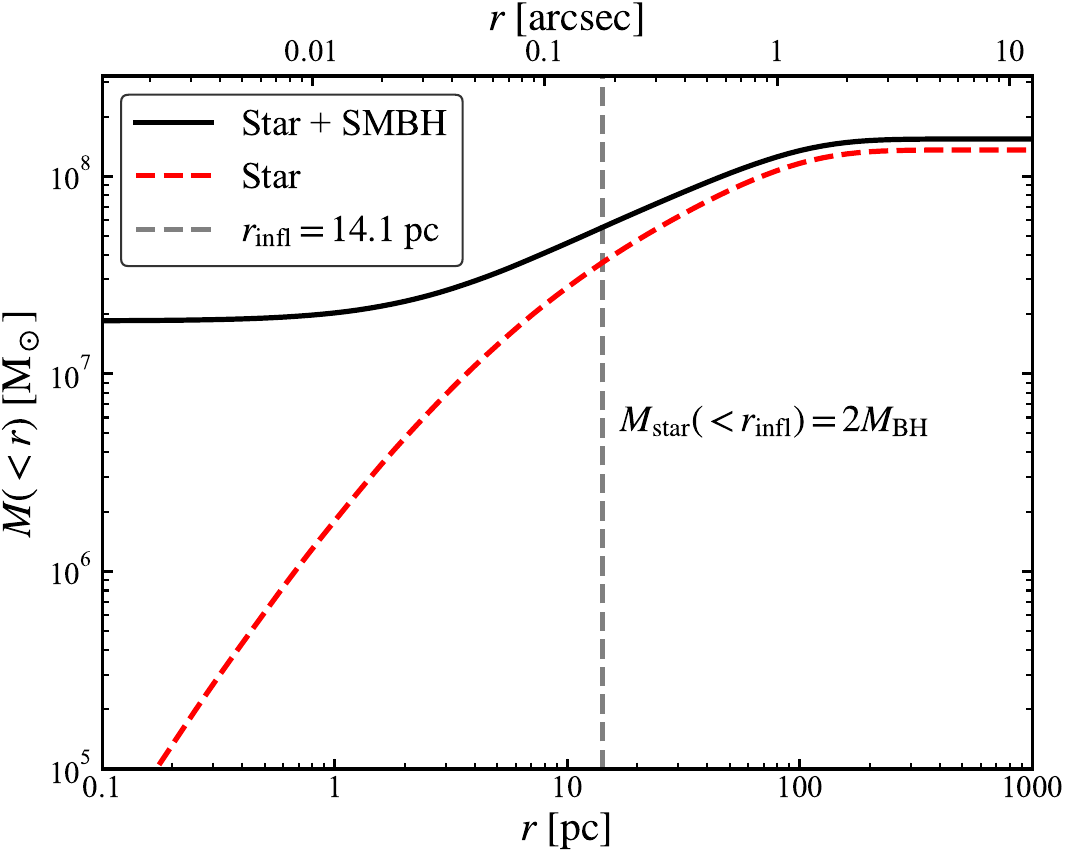}
	\caption{Enclosed mass profile of the dynamical model of M60-UCD1. The black solid line denotes the total mass while the red dashed line denotes the stellar component. The grey dashed line indicates the sphere of influence of the SMBH with $r_{\rm infl}=14.1~\rm pc$.\label{fig:mass_profile}}
\end{figure}

First, a three-dimensional triaxial mass distribution is obtained by deprojecting the observed surface brightness.
The observed surface brightness $\Sigma(R^\prime, \theta^\prime)$ is parameterized using the multi-Gaussian expansion (MGE), expressed as a sum of two-dimensional Gaussian components: 
$$\Sigma(R^\prime, \theta^\prime)=\sum_{j=1}^N \frac{L_j}{2\pi\sigma^{\prime2}_jq^\prime_{j}}\exp[-\frac{1}{2\sigma^{\prime2}}(x^{\prime2}_j+\frac{y^{\prime2}_j}{q^{\prime2}_j})],$$
where $(R^\prime, \theta^\prime)$ is the polar coordinate on the sky plane, $N$ is the number of Gaussian component, and the subscript $j$ indicates the $j$-th Gaussian. 
For each Gaussian with a luminosity of $L_j$, $(x^{\prime}_j, y^{\prime}_j)$ is the coordinate at its coordinate system with a specific position angle, $q^\prime_j$ is the axis ratio, and $\sigma^\prime_j$ is the dispersion along the major axis.
By assuming that the intrinsic density is triaxial, the three-dimensional stellar density distribution can be written as
\begin{align*}
    \rho(x,y,z)=& \sum_{j=1}^N (M_\ast/L)\frac{L_j}{(\sqrt{2\pi}\sigma_j)^3p_{j}q_j}\\
    &\times \exp[-\frac{1}{2\sigma_j^{2}}(x^{2}+\frac{y^{2}}{p^{2}_j}+\frac{z^2}{q^2_j})],
\end{align*}
where the mass-to-light ratio $M_\ast/L$ is a free parameter, and $p_j=b_j/a_j$ and $q_j=c_j/a_j$ are the axis ratios for the ellipsoid.
For modeling M60-UCD1, we adopt MGE parameters from \citet{2014Natur.513..398S} based on \emph{Hubble Space Telescope}/F475W ($g$-band) observations, and fixed axis ratios of $b/a=0.99$ and $c/a=0.7$.
Dark matter is neglected for the extreme compactness of M60-UCD1.
Given $M_\ast/L$ and the mass of the central black hole $M_{\rm BH}$, one set of gravitational potential of the galaxy is determined.

Second, the Schwarzschild method will derive a numerical solution of the galaxy's dynamics, by assigning weights for a set of orbits to fit the observed stellar kinematics.
The orbit library samples the three integrals of motion $(E, I_2, I_3)$ on a $(36, 9, 8)$ grid, yielding a total of $3\times36\times9\times8=7776$ orbits without dithering.
As for the stellar kinematics, we employ a kinematics map \citep[][private communication]{2014Natur.513..398S} that is adaptively binned with the Voronoi binning method, based on Gemini/NIFS observations with a spatial pixel size of 0\farcs05 and a resolving power of $R\sim5000$.

Finally, we explore a parameter grid consisting of 27 $M_\ast/L$ values, linearly sampled between 2.4 and 5.0, and 15 $\log( M_{\rm BH}/M_\odot)$ values, linearly sampled between 6.4 and 7.8, to identify the best-fit dynamical model that minimizes the $\chi^2$ value.
The best-fit model yields $M_\ast/L=4.2$ corresponding to a total stellar mass of $1.36\times10^{8}~\rm M_\odot$, and a black hole mass of $1.85\times10^{7}~\rm M_\odot$, which serves as the basic setup for the gravitational potential of our simulations in Section \ref{sec:simulation}.
This configuration is well consistent with the dynamical modeling in \citet{2014Natur.513..398S} with $M_\ast=(1.2\pm0.4)\times10^{8}~\rm M_\odot$ and $M_{\rm BH}=2.1_{-0.7}^{+0.4}\times10^{7}~\rm M_\odot$.

\section{Simulation Setup}\label{sec:simulation}

\subsection{Governing Equations}\label{sec:equation}
We perform a suite of three-dimensional hydrodynamical simulations with the Godunov-type, grid-based code {\sc pluto}\footnote{\url{https://plutocode.ph.unito.it/}} \citep[version 4.4patch2;][]{2007ApJS..170..228M}.
The numerical setup basically follows \citet{2025ApJ...988...68S}, which introduced the method in detail.

In brief, the equations we solve are
\begin{eqnarray*}
\frac{\partial\rho}{\partial t} + \nabla\cdot(\rho \bm{v}) &= \dot{\rho}_w, \\
\frac{\partial\rho \bm{v}}{\partial t} + \nabla\cdot(\rho \bm{vv}+p\mathbb{I}) &= -\rho\nabla\Phi+\dot{\bm{m}}_w, \\
\frac{\partial E_t}{\partial t} + \nabla\cdot[(E_t+p)\bm{v}] &= -\rho\bm{v}\cdot\nabla\Phi + \dot{\rho}_w\Phi +\dot{E}_w+\dot{Q},
\end{eqnarray*}
where $\rho$ is the mass density, $p$ is the thermal pressure, $\bm{v}$ is the velocity, $E_t=p/(\gamma-1)+1/2\rho v^2$ is the total energy density, $\mathbb{I}$ is the identity matrix, $\gamma=5/3$ is the adiabatic index for an ideal equation of state.
The gravitational potential $\Phi$ comprises the central SMBH and the stellar component (Figure \ref{fig:mass_profile}), which is derived from the dynamical modeling of M60-UCD1 in Section \ref{sec:dynamics}.
Stellar winds are implemented by the source terms of mass $\dot{\rho}_w$, momentum $\dot{\bm{m}}_w$, and energy $\dot{E}_w$.
$\dot{Q}$ is the term for radiative cooling, for which we adopt a cooling function generated by \textsc{Cloudy} \citep[version C23;][]{2023RMxAA..59..327C} for an optically thin plasma with a solar abundance.
To circumvent the time-step limitation arising from short cooling timescales in some regions, we adopt the exact integration cooling scheme proposed by \citet{2009ApJS..181..391T}.
Magnetic fields are neglected for simplicity and because their strength is poorly constrained.
Although magnetic fields would be important close to the black hole horizon, the gas configuration of weakly magnetized stellar winds at a larger scale that is the focus of this work, could be primarily governed by hydrodynamics. 
We discuss the role of magnetic fields and the possible limitation for neglecting them in Section \ref{subsubsec:bfield}.

\subsection{Grid and Run Setup} \label{sec:grid}

To alleviate the severe time-step in small scales, the simulation is divided into four separate runs which start with the coarsest grid and successively increase resolution in subsequent runs.
The simulation domain has a volume of $400^3~\rm pc^3$ to encompass the bulk of M60-UCD1.
The initial run adopts a $224^3$ nested Cartesian grid with 6 refinement levels, where the smallest pixel  corresponds to $1/2^6$ of the box length.
For the subsequent run, the simulation is initialized with the last snapshot of the previous run, adding three additional refinement levels to achieve a resolution $2^3$ times higher in the vicinity of the SMBH.
Consequently, the final run employs a $512^3$ grid with 15 refinement levels, achieving a maximum resolution of $3.8\times10^{-4}~\rm pc=431~r_{\rm g}$ where $r_{\rm g}$ denotes the gravitational radius of the SMBH.
The SMBH is positioned at the center of the grid, $(0, 0, 0)$. 
The $xyz$ axes are aligned with the major, intermediate, and minor axis of the triaxial model of M60-UCD1, as described in Section \ref{sec:dynamics}.

From the coarsest to the finest run, the simulation times are $5.0\times10^6$, $2.2\times10^5$, $4.2\times10^4$, and $3.8\times10^3$ years, respectively, resulting in a total simulation time of 5.3 Myr.
A quasi-steady state for the BH accretion is observed in each run.

\begin{figure}[hbpt!]
	\centering
	\includegraphics[width=0.48\textwidth]{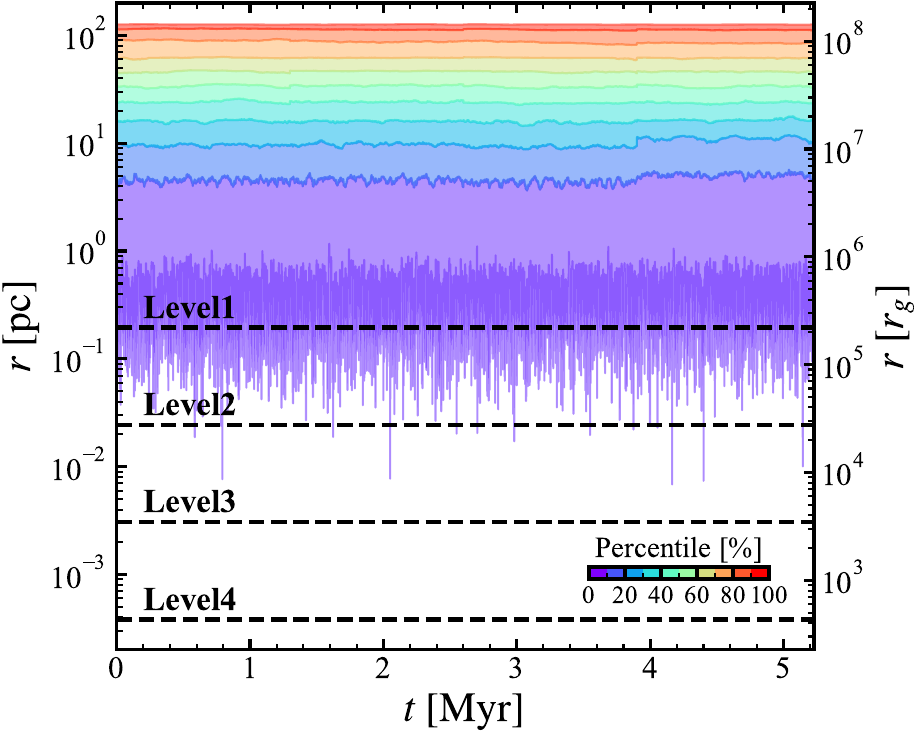}
	\caption{Radius percentiles of the sampled AGB stars versus time in the simulation. For comparison, the resolutions of different runs (Section \ref{sec:grid}) are indicated by the dashed horizontal lines, with ``Level1" representing the coarsest run and ``Level4" representing the finest run. The discontinuity at 3.9 Myr is due to the random sampling of AGB stars. \label{fig:sample_percentile}}
\end{figure}

\subsection{Stellar winds and orbits}\label{sec:wind}

Similar to \citet{2025ApJ...988...68S}, we assume that stellar winds from the old stellar population are dominated by a minor population of asymptotic giant branch (AGB) stars during the thermally pulsating (TP-AGB) phase.
To estimate the mass-loss from the stars, we refer to the MESA Isochrones and Stellar Tracks \citep[MIST,][]{2016ApJS..222....8D,2016ApJ...823..102C} to predict the number and average mass-loss rate of TP-AGB stars.
We adopt the Kroupa canonical initial mass function \citep{2001MNRAS.322..231K, 2002Sci...295...82K} and assume a solar metallicity and a stellar age of 12 Gyr, which is consistent with the old stellar population of solar metallicity in M60-UCD1 as measured by \citet{2013ApJ...775L...6S} using optical spectroscopy.
For a total stellar mass of $1.36\times10^8~\rm M_\odot$ and a simple stellar population with an age of 12 Gyr, the isochrones predict the presence of 1478 TP-AGB stars and a total mass-loss rate of $4.3\times10^{-4}~\rm M_\odot~yr^{-1}$.
The time-averaged mass-loss rate for individual TP-AGB stars is thus $\dot{M}_{\rm w}=2.9\times10^{-7}~\rm M_\odot~yr^{-1}$. 
We assume a constant mass-loss rate and neglect the contribution from red giant branch stars as in \citet{2025ApJ...988...68S}.
Each AGB star will lose $\Delta M=0.38~\rm M_\odot$ according to the white dwarf initial-final mass relation from \citet{2018ApJ...866...21C}, assuming all the mass-loss happens during the AGB phase.
Therefore, the mass-loss time scale, i.e., the lifetime of these AGB stars in the simulation is determined to be $\Delta M/\dot{M}_{\rm w}=1.30~\rm Myr$. 

Although the chemical composition is expected to change during the AGB phase, we assume that AGB winds have the same metallicity as the underlying stellar population for simplicity.
This assumption is justified by that low-mass AGB stars with initial masses of $\approx1~M_\odot$ at solar metallicity are predicted to experience only modest metal enrichment ($\Delta Z/Z_{\rm ini}<0.1$), owing to inefficient or absent third dredge up \citep[e.g.,][]{2010MNRAS.403.1413K,2015ApJS..219...40C,2016ApJ...825...26K,2018MNRAS.475.2282V}.

For the motion of individual AGB stars, we first randomly sample particles from the trajectories of the orbit library with solved orbit weights in Section \ref{sec:dynamics}.
The sampling decides initial positions and velocities of the stars, and their motions are then integrated on-the-fly in the simulation.
Once the time elapsed reaches the lifetime of 1.30 Myr, the present AGB stars are instantaneously replaced by a new sample.

The time evolution of the radius percentiles of sampled AGB stars is shown in Figure \ref{fig:sample_percentile}. 
The apparent discontinuity at 3.9 Myr present in Figure \ref{fig:sample_percentile} arises from the random sampling of AGB stars, and is found to have a negligible impact on the overall simulation results.
For each star, we adopt the same wind injection prescription as \citet{2025ApJ...988...68S}: wind material is injected within a sphere of 4 cells, and has a temperature of 3000 K.
The velocity is the combination of the instantaneous orbital velocity and an isotropic wind velocity of $10~\rm km~s^{-1}$. 

We note that, in reality, AGB stars would enter and evolve off the AGB phase individually rather than being replaced simultaneously as assumed in our simulations.
We test the potential impact of gradual replacement and find that it leads to differences of only about 10\% in SMBH accretion rate at the coarsest level.
The little difference is owing to the large number of AGB stars and their co-rotating dynamics, indicating that the present wind-injection prescription can reasonably capture the collective  behavior of stellar winds.

\subsection{Ambient Environment}\label{sec:ram}
M60-UCD1 is located at the outskirt of the galaxy M60, a massive elliptical member of the Virgo cluster, thus the ambient environment could have a significant impact.
Two types of medium, the intracluster medium (ICM) of the Virgo cluster and the interstellar medium (ISM) of M60, can serve as external gas inflows, potentially affecting the wind accretion by ram pressure stripping.
In addition to the {\tt Fiducial} run, we include these media separately in two simulations:
\begin{enumerate}
	\item \texttt{ICM}. M60 is suggested to be undergoing infall toward the cluster's central galaxy, M87. The infall through the ICM and the consequent gas-stripping are manifested by the X-ray surface brightness discontinuity in the direction toward M87 \citep{2017ApJ...847...79W}. As a satellite potentially bound to M60 \citep{2014Natur.513..398S}, M60-UCD1 may also experience the ram pressure stripping from the Virgo ICM. In the simulation, we adopt an infall velocity $v_{\rm ICM}$ of $1030~\rm km~s^{-1}$ and the ICM surrounding M60 has a number density $n_{\rm ICM}$ of $5.6\times10^{-5}~\rm cm^{-3}$ and a temperature $kT_{\rm ICM}$ of $1.37~\rm keV$, according to the X-ray spectral fitting by \citet{2017ApJ...847...79W}. For simplicity, we ignore M60-UCD1's motion relative to M60 due to the large uncertainty and the low relative velocity ($\sim 250~\rm km~s^{-1}$) compared to the infall velocity. The ICM inflow is assumed to have an inclination angle of $\approx60^\circ$, based on the direction toward M87, transformed into the coordinate system of the simulation.

	\item \texttt{ISM}. M60-UCD1 has a projected distance of 7 kpc from the center of M60. While the exact line-of-sight distance from M60 is unknown, M60-UCD1 should have closely interacted with M60 to lead to its extreme compactness, and \citet{2014Natur.513..398S} suggested that an orbit with a pericenter of 1 kpc and an apocenter of 30 kpc could favor the formation of M60-UCD1. Therefore, it is possible that M60-UCD1 is embedded within M60's hot ISM, which can shield M60-UCD1 from the ram pressure of the ICM. For the ISM case, we set the ISM at the distance of M60-UCD1 to $n_{\rm ISM}=4.9\times10^{-3}~\rm cm^{-3}$ and $kT_{\rm ISM}=0.9~\rm keV$,  according to the ISM radial profile of M60 \citep[][Figure 7 therein]{2014ApJ...787..134P}. The velocity between the ISM and M60-UCD1 is set to $250~\rm km~s^{-1}$, consistent with the velocity dispersion at the projected distance of M60-UCD1 from M60 \citep{2008ApJ...674..869H}. Since its motion relative to M60 is still unclear, the inclination of the ISM inflow is set to the same value of the ICM inflow for simplicity. In spite of the uncertainty, it has been found that the gas loss during ram pressure stripping is insensitive to the inclination angle as long as the galaxy is not moving edge-on \citep{2000Sci...288.1617Q,2006MNRAS.369..567R, 2009A&A...500..693J}. 
\end{enumerate}

We summarize the properties of the ambient environment in Table \ref{tab:sum}. 
We note that the equivalent {\it Bondi} accretion rates for the ICM and ISM inflows, estimated to be $4\times10^{-11}~M_\odot~\rm yr^{-1}$ and $4\times10^{-8}~M_\odot~\rm yr^{-1}$, respectively, are much lower than the mass-loss rate of the stellar winds. 
Therefore, direct accretion onto the SMBH from the ambient medium is negligible compared to the stellar winds (also see Section \ref{subsec:xray}). 
For the numerical treatment, the medium is injected from the boundary of the simulation box with specific density, temperature, and velocity vector.

\begin{figure*}[hbpt!]
	\centering
	\includegraphics[width=1.0\textwidth]{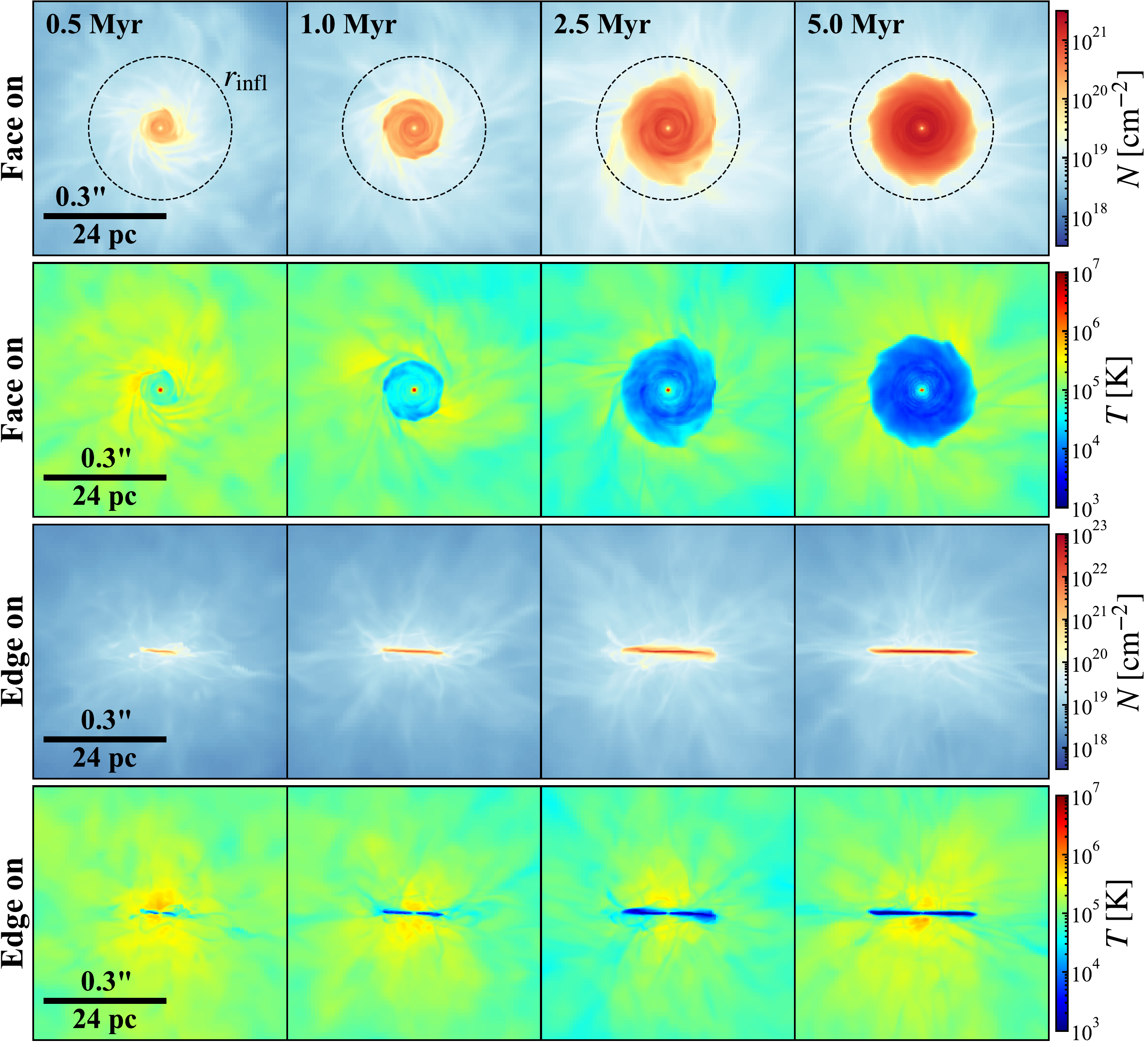}
	\caption{Projected gas density and temperature of the {\tt Fiducial} simulation. The top and bottom two rows represent the face-on and edge-on view, respectively. From left to right are for snapshots at 0.5, 1.0, 2.5, and 5.0 Myr. The dashed black circles mark the gravitational sphere of influence of the SMBH. Each panel has a length of 50 pc. The black solid lines indicate a spatial scale of 24 pc and an angular scale of 0.3 arcsec at the distance of M60-UCD1 \citep[16.5 Mpc;][]{2009ApJ...694..556B}, corresponding to the observed half-light radius of M60-UCD1 \citep{2013ApJ...775L...6S}. The gas disk has a clockwise rotation in the face-on view. \label{fig:fiducial_density_temperature}}	
\end{figure*}

\subsection{Boundary and Initial Conditions}\label{sec:conditions}
The boundary condition for the {\tt Fiducial} simulation is ``outflow"  while it is set to be the external inflow described in Section \ref{sec:ram} for the {\tt ICM/ISM} simulations.  
As in \citet{2025ApJ...988...68S}, the accretion onto the SMBH is mimicked by ``removing'' the gas in the central $2^3$ cells, which corresponds to an effective accretion radius of $r_{\rm acc}=431~r_{\rm g}$ at the finest level.

The initial condition for the {\tt Fiducial} simulation is a uniformly distributed gas with a number density of $10^{-4}~\rm cm^{-3}$ and a temperature of $10^4~\rm K$.
For the {\tt ICM/ISM} simulation, the simulation is performed first without stellar winds injection for 2.0 Myr to reach a steady state and the final snapshot is used as the initial condition.
We note that the low-density gas at the start will be rapidly replaced by the stellar winds in the {\tt Fiducial} simulation or the high-speed inflowing medium in the {\tt ICM/ISM} simulation.

\section{Simulation results}\label{sec:results}
\subsection{Corona-Disk-Halo Configuration}
\begin{figure}[hbpt!]
	\centering
	\includegraphics[width=0.48\textwidth]{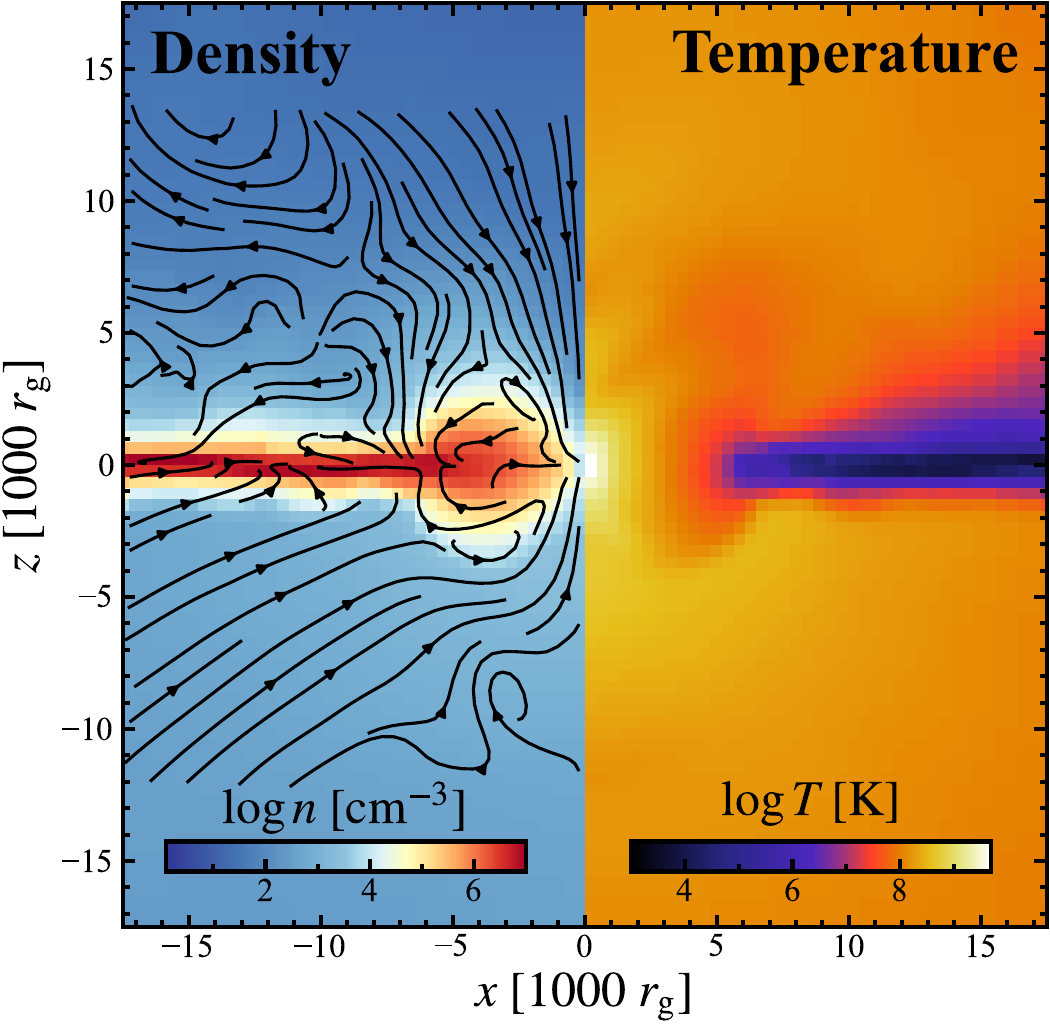}
	\caption{A zoom-in view of the {\tt Fiducial} simulation at 5.3 Myr, the last snapshot of the finest run. The left half  shows the azimuthally-averaged gas density with superposed streamlines indicating the fluid velocity, and the right half shows the azimuthally-averaged gas temperature. \label{fig:fiducial_zoomin}}
\end{figure}

As shown in Figure \ref{fig:fiducial_density_temperature}, the slow stellar winds injected from about 1500 AGB stars accumulate within the galaxy rather than escaping.
Due to the non-zero net angular momentum of the oblate galaxy, which is characterized by a significant fraction of stars on co-rotating orbits \citep{2014Natur.513..398S}, the majority of the wind material eventually forms a cold gaseous disk ($\lesssim10^4~\rm K$) within the sphere of influence of the SMBH, which has a radius of $\sim$14 pc.
At 5 Myr, the accumulated mass of the gas disk is about $1400~\rm M_\odot$ in the {\tt Fiducial} simulation. 
Surrounding the gas disk is a tenuous, warm halo ($\sim 10^4-10^7~\rm K$), consisting of shock-heated winds that have not undergone effective radiative cooling.
The cold, thin disk supplies accretion onto the SMBH and transforms into a hot, geometrically thick corona ($\gtrsim10^7~\rm K$) in the immediate vicinity of the SMBH ($<10^4~r_{\rm g}$), as shown in Figure \ref{fig:fiducial_zoomin}.

\begin{figure}[hbpt!]
	\centering
	\includegraphics[width=0.48\textwidth]{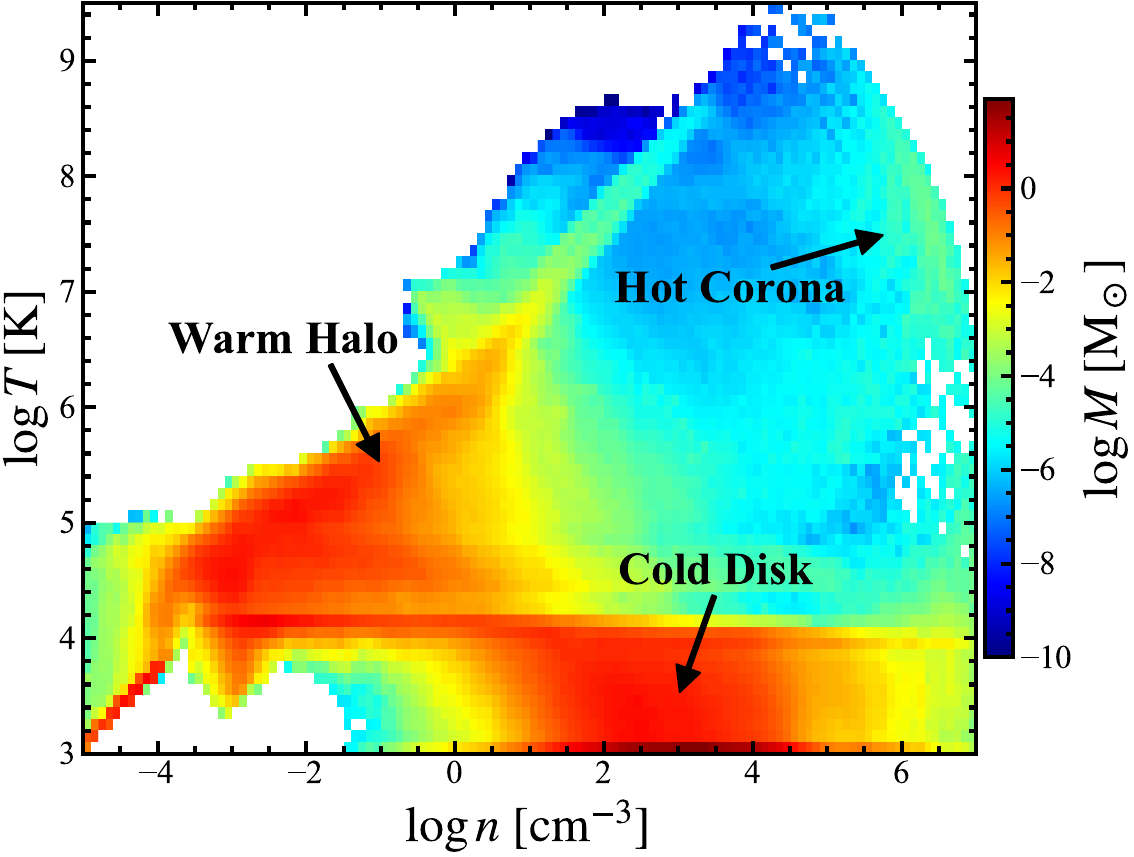}	
	\caption{Temperature versus density distribution of the {\tt Fiducial} simulation at 5.3 Myr. The color bar represents the mass occupation of the gas. Three dominant components (hot corona, cold disk, and warm halo) are indicated by the arrows. \label{fig:fiducial_hist}}
\end{figure}

\begin{figure*}[hbpt!]
	\centering
	\includegraphics[width=1.0\textwidth]{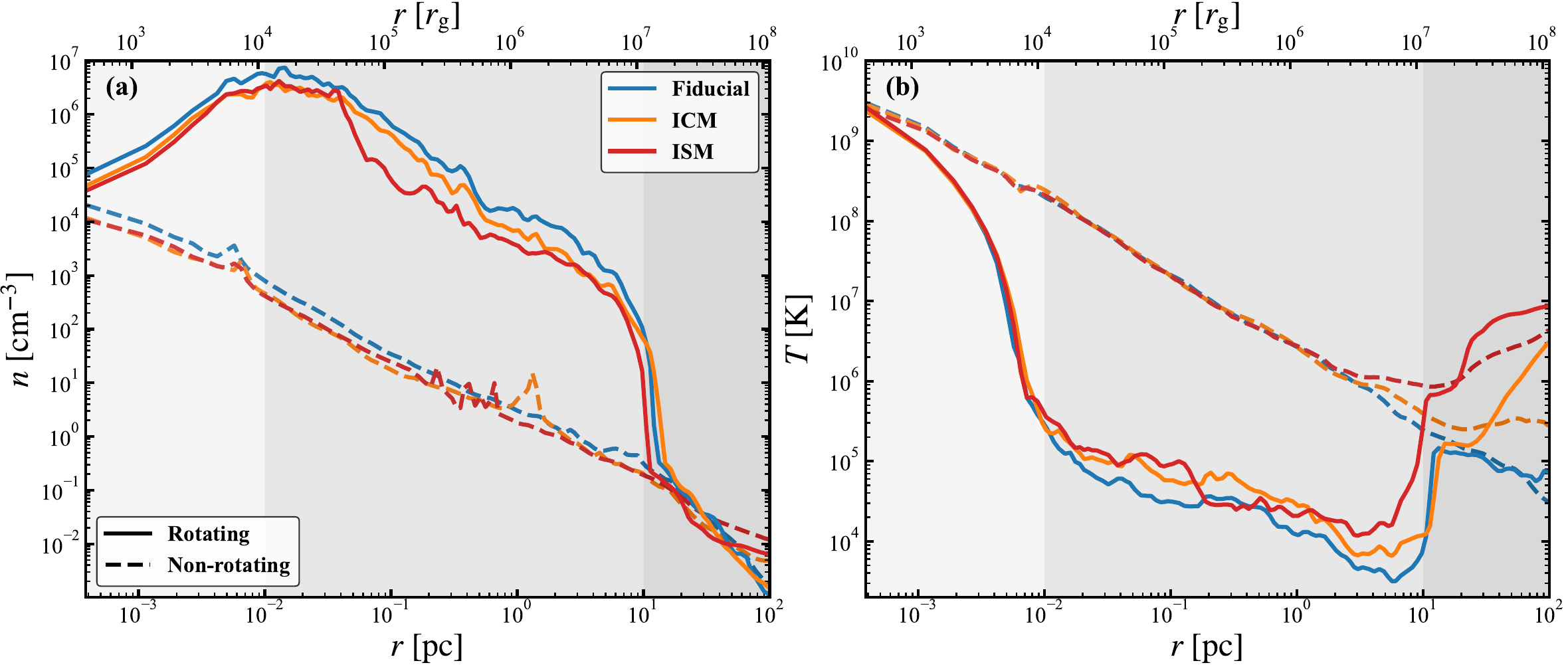}
	\caption{Mass-weighted density (left) and temperature (right) radial profiles. The ``rotating" (sold curves) component indicates rotation-dominated gas such as the cold disk and the hot corona, while the ``non-rotating" (dashed curves) component indicates pressure-supported gas such as the thermalized halo. These two components are defined by the ratio of tangential velocity of the gas to the circular velocity at that radius, with a critical value of 0.5. The shaded regions in both panels indicate the characteristic radius ranges dominated by the hot corona ($<10^{-2}~\rm pc$), the cold disk ($10^{-2}-10~\rm pc$), and the warm halo ($>10~\rm pc$), respectively. \label{fig:profile}}
\end{figure*}

These three components are clearly distinguished in the temperature-density diagram (Figure \ref{fig:fiducial_hist}): (i) the gas disk has a temperature below $2\times10^4$~K and a density of $10^2-10^6~\rm cm^{-3}$ increasing from the outer to the inner regions; (ii) the warm halo has a density of $\sim10^{-4}-100~\rm cm^{-3}$ and a temperature up to $\sim 10^{7}~\rm K$, which is correlated with the density; (iii) the innermost hot corona, although contributing only a small fraction of the gas, has significantly high density and temperature, which dominates the total X-ray emission produced by the system (Section~\ref{subsec:xray}).
This three-component configuration (see Figures \ref{fig:fiducial_density_temperature} and \ref{fig:profile}) bears qualitative resemblance to the gas distribution observed in the vicinity of Sgr A*, where a cool ($10^4~\rm K$) gaseous disk is embedded within a surrounding hot plasma of $\sim10^7~\rm K$ \citep{2019Natur.570...83M}.
From a theoretical perspective, a similar multiphase structure has also emerged in wind-fed simulations for M31* \citep{2025ApJ...988...68S} and Sgr A* \citep{2020ApJ...888L...2C, 2024ApJ...974...99B, 2025A&A...693A.180C}, which hints on the generic behavior of weakly accreting SMBHs fed by stellar winds.
In Figure \ref{fig:profile}, we present the radial distributions of gas density and temperature.
The cold disk ($10^{-2}-10~\rm pc$) increases in density toward the center until it transitions into the hot corona at approximately $10^{-2}~\rm pc$, within which the temperature dramatically rises to X-ray-emitting levels.
For the warm halo, both density and temperature increase as the radius decreases, and the temperature follows $T\propto r^{-1}$, indicating thermalization and inefficient radiative cooling of the gas.
In the {\tt ICM/ISM} simulations, the halo-dominated region ($\gtrsim10~\rm pc$) also contains the inflowing hot medium and the stripped tail, as reflected by the increasing temperature at larger distances from the SMBH.  

\begin{deluxetable*}{@{}ccccccc@{}}
 	\tablecolumns{7}
 	\tablecaption{Summary on simulation setup and results\label{tab:sum}}
 	\tablehead{\colhead{Simulation} & \colhead{External inflow} & $n_{\rm inflow}$          & $kT_{\rm inflow}$     & $v_{\rm inflow}$            & \colhead{$\dot{M}_{\rm acc}$} & \colhead{$L_{\rm X}$}\\
 	                         &                           & \colhead{($\rm cm^{-3}$)} & \colhead{($\rm keV$)} & \colhead{($\rm km~s^{-1}$)} & \colhead{(${\rm M_\odot~yr^{-1}}$)}&\colhead{(${\rm erg~s^{-1}}$)}\\
 	           \colhead{(1)} & \colhead{(2)} & \colhead{(3)} & \colhead{(4)} & \colhead{(5)} & \colhead{(6)} & \colhead{(7)}           
 	           }
 	\startdata
 	{\tt Fiducial} & No  &    -                &   -    &    -    & $1.6\times10^{-5}$ & $7.4\times10^{37}$\\
 	{\tt ICM}      & ICM & $5.6\times10^{-5}$ & $1.4$ & $1030$ & $8.3\times10^{-6}$ & $3.9\times10^{37}$ \\
 	{\tt ISM}      & ISM & $4.9\times10^{-3}$ & $0.9$ & $250$  & $7.5\times10^{-6}$ & $3.2\times10^{37}$
 	\enddata
 	\tablecomments{(1) Name of the simulation; (2) Type of the external inflow; (3)-(5) Density, temperature, and velocity of the inflow; (6)-(7) Average accretion rate and 0.3--8 keV X-ray luminosity over the last 1000 yr.}
 \end{deluxetable*}

\subsection{Impact of ICM and ISM}
\begin{figure*}[hbpt!]
	\centering
	\includegraphics[width=1.0\textwidth]{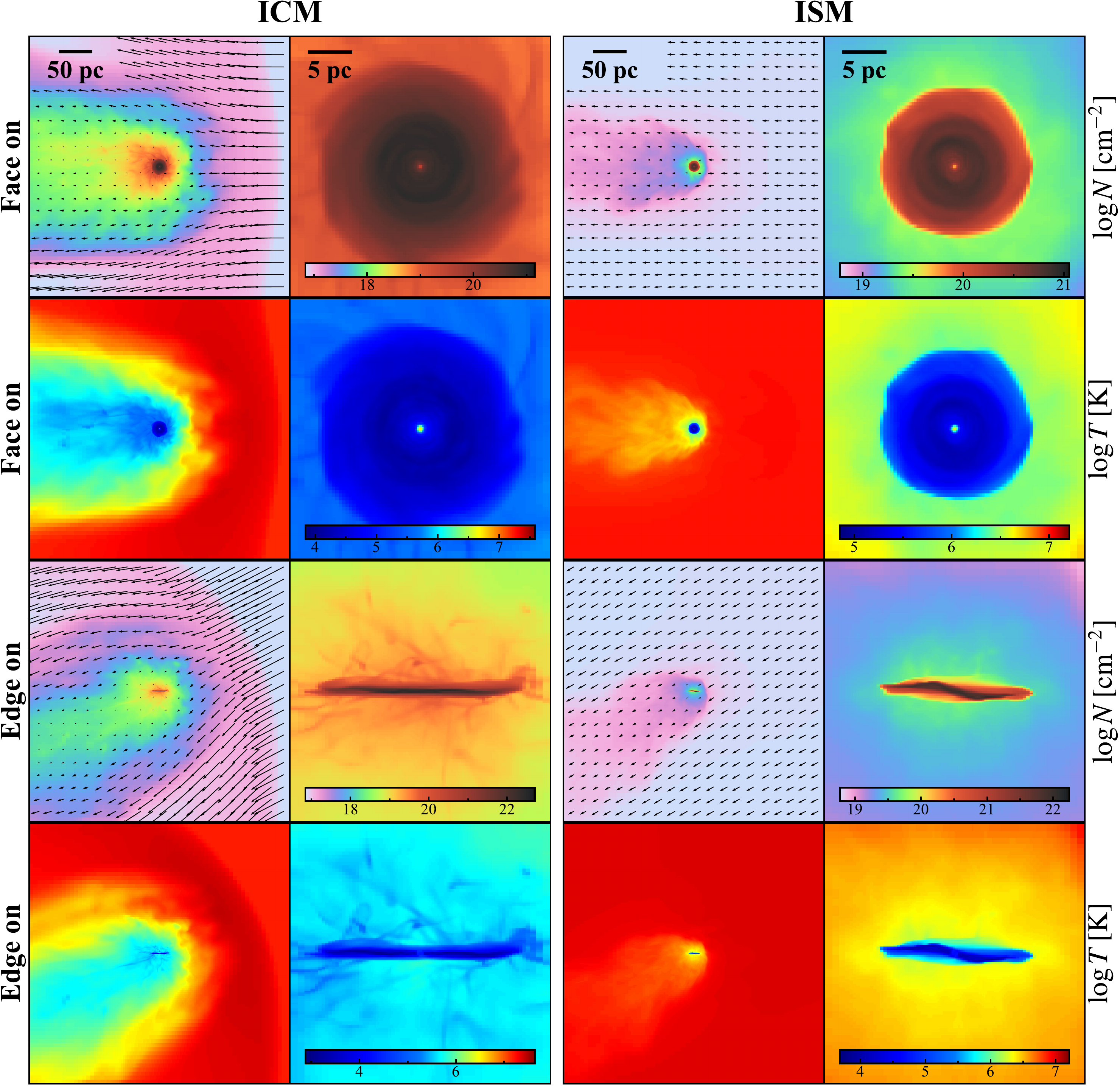}
	\caption{Projected gas density and temperature of the {\tt ICM} (left two panels) and {\tt ISM} (right two panels) simulations at 5.0 Myr. For each simulation, the left and right panels represent the large-scale (400 pc) and small-scale (30 pc) distributions, respectively. Similar to Figure \ref{fig:fiducial_density_temperature}, the top two rows represent the face-on projected density and temperature while the bottom two rows represent the edge-on view. Arrows indicate the fluid velocity. }
    \label{fig:icmism}
\end{figure*}

\begin{figure}[hbpt!]
	\centering
	\includegraphics[width=0.48\textwidth]{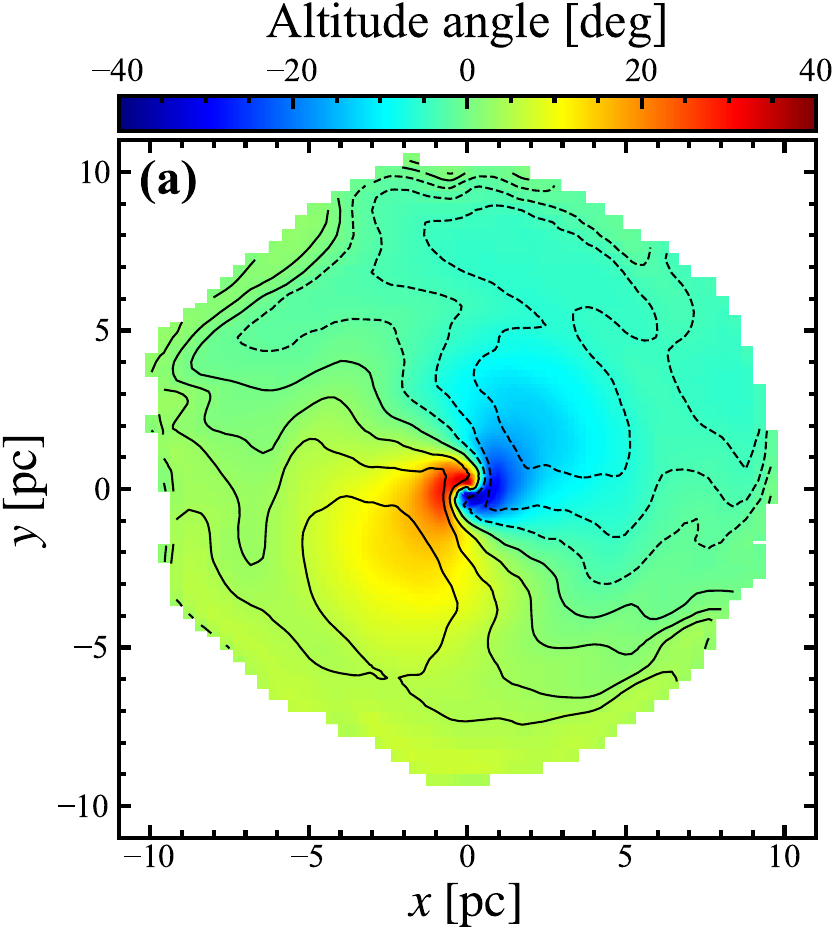}
	\includegraphics[width=0.48\textwidth]{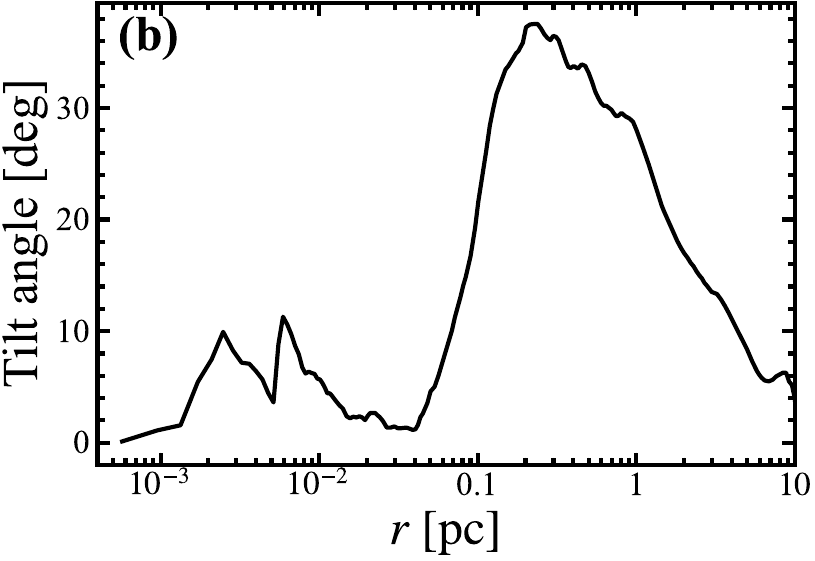}
	\caption{(a): distribution of the altitude angle of the warped gas disk in the {\tt ISM} simulation. Here, the altitude angle is define as $\arctan(\bar{z}/r)$ where $\bar{z}$ is mass-weighted coordinate along $z$-axis, i.e., the axis perpendicular to the disk, and $r$ is the cylindrical radius. The contours display the distribution of $\bar{z}$ at 0.2 pc intervals, with solid lines representing $\bar{z}\ge0~\rm pc$ and dashed lines representing $\bar{z}<0~\rm pc$. (b): tilt angle versus cylindrical radius. The tilt angle is $(\theta_{\rm max}-\theta_{\rm min})/2$ where $\theta_{\rm max}$ and $\theta_{\rm min}$ are maximum and minimum of the altitude angle at that radius, respectively. \label{fig:disk_tilt}}
\end{figure}

Galaxies moving through a hot ambient medium can lose a substantial amount of their gas due to ram pressure stripping.
As shown in Figure~\ref{fig:icmism}, a portion of the wind material is removed by the ram pressure and the stripped gas forms a tail aligned with the direction of the inflowing gas, i.e., the opposite direction of the galaxy's motion.
Whether subjected to ram pressure from the ICM or ISM, a gas disk eventually survives, with masses of $720~M_\odot$ and $350~M_\odot$ in the {\tt ICM} and {\tt ISM} simulations, respectively, which are about 50\% and 25\% of that in the {\tt Fiducial} simulation.
Particularly for the {\tt ISM} case where the ram pressure is stronger, the gas disk becomes warped due to the torque from the inclined inflow (see the zoomed edge-on view of the {\tt ISM} simulation in Figure \ref{fig:icmism}).
The tilt angle reaches up to $40^\circ$ at 0.2 pc and decreases to nearly zero with increasing radius (see Figure \ref{fig:disk_tilt}).
The warped gas disk is also in agreement with the findings of \citet{2014MNRAS.440L..21H}, who showed that an inclined flow ($45^\circ$) leads to a S-shaped (integral-shaped) warped disk while a perpendicular flow leads to a U-shaped (bow-shaped) warp \citep{1998A&A...337....9R}.

The significantly different disk masses among these three simulations may provide valuable observational insights for the properties of the external environment.
However, we note that the presence of supernovae feedback, AGN activity, and tidal disruption events could substantially affect the steady state of the gas disk on longer time scales.
These effects will introduce uncertainties when attempting to infer stellar wind properties and the external environments via the accumulated gas disk.

\subsection{Accretion Rate and X-ray Emission}
\label{subsec:xray}
The cold gas disk feeds the accretion and the stellar wind material is heated due to gravity, forming the innermost hot corona and emitting substantial X-ray emission.
To distinguish whether the X-ray counterpart of M60-UCD1 could be the accretion signature of the SMBH, we calculate the synthetic X-ray luminosity following the method described in \citet{2025ApJ...988...68S}.
Briefly, we account for thermal emission from the hot plasma assuming collisional ionization equilibrium and a solar metallicity.
The thermal spectra, as a function of temperature, are extracted from ATOMDB\footnote{\url{http://www.atomdb.org}}. 
The collective spectra of all the gas within the simulation domain are then integrated over 0.3--8 keV to derive the broadband X-ray luminosity.

\begin{figure}[hbpt!]
	\centering
	\includegraphics[width=0.5\textwidth]{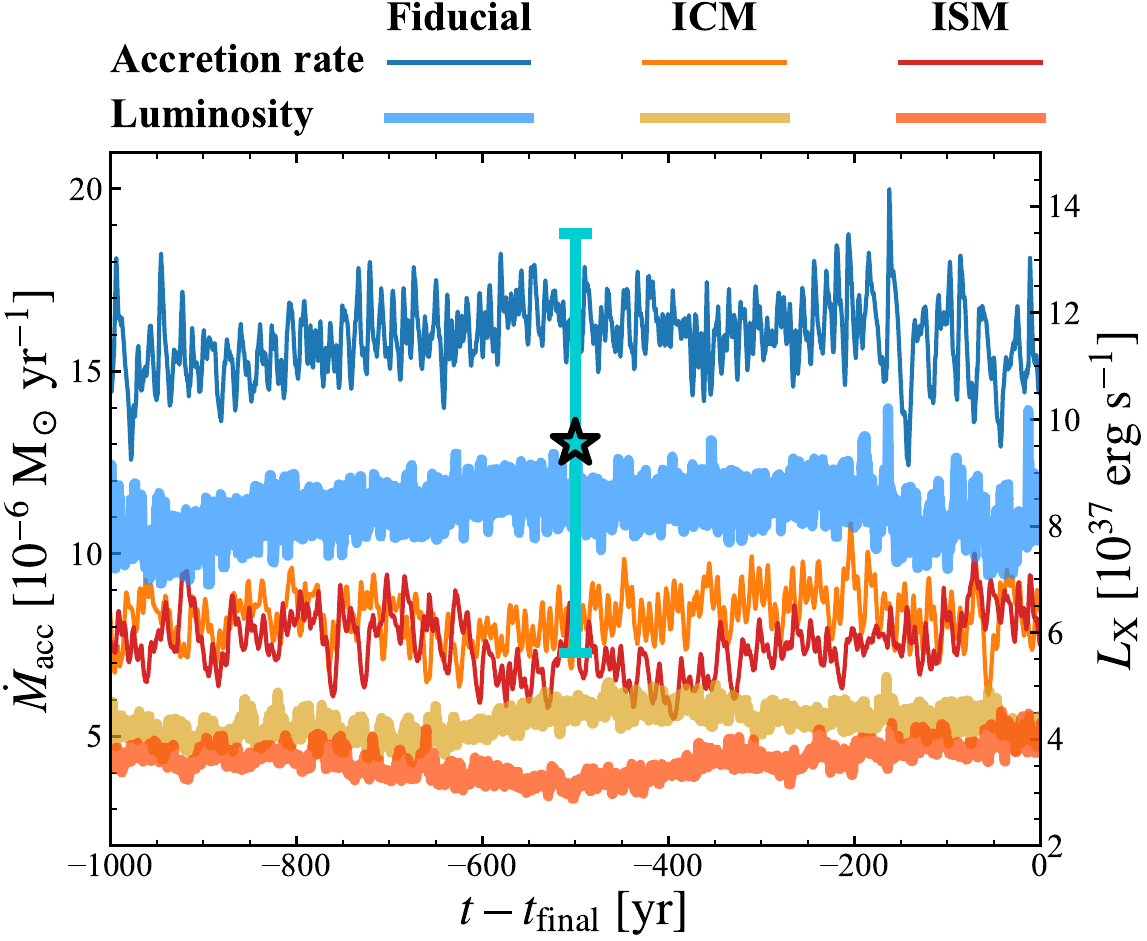}
	\caption{Evolution of the accretion rate and the synthetic X-ray luminosity over the last 1000 yr at the quasi-steady state. The thin and thick lines represent the accretion rate and X-ray luminosity, respectively. For the purpose of comparison, the star symbol indicates the {\it Chandra} X-ray counterpart with a 0.3--8 keV luminosity of $9.5\times10^{37}~\rm erg~s^{-1}$ and the error bar shows the long-term variability ranging from $5.6\times10^{37}~\rm erg~s^{-1}$ to $1.3\times10^{38}~\rm erg~s^{-1}$ \citep{2013ApJS..204...14L}. 
    \label{fig:accrate_lx}}
\end{figure}

Figure \ref{fig:accrate_lx} shows the evolution of the accretion rate and the synthetic X-ray luminosity at the quasi-steady state of the finest run.
With an effective accretion radius of $430~r_{\rm g}$, the average accretion rates are $1.6\times10^{-5}~\rm M_\odot~yr^{-1}$, $8.3\times10^{-6}~\rm M_\odot~yr^{-1}$, and $7.5\times10^{-6}~\rm M_\odot~yr^{-1}$, for the {\tt Fiducial}, {\tt ICM}, and {\tt ISM} simulations, respectively.
These correspond to $(2-4)\times10^{-5}$ of the Eddington rate of the SMBH, which indicates a low-accretion-rate state in agreement with the estimate by \citet{2013ApJ...775L...6S}.
As for the X-ray luminosity, we derive $7.4\times10^{37}~\rm erg~s^{-1}$, $3.9\times10^{37}~\rm erg~s^{-1}$, and $3.2\times10^{37}~\rm erg~s^{-1}$ for the three simulations, respectively.
The innermost hot corona dominates the X-ray emission (see the inset of Figure \ref{fig:synthetic_spectra}), while the tenuous halo and the inflowing ICM/ISM contribute negligibly, accounting for less than $<1\%$.
Both the accretion rates and X-ray luminosities exhibit modest variations of $<10\%$ across all simulations and those of the {\tt ISM} simulation shows the strongest fluctuation, which may originate from turbulence induced by the high ram pressure.
These results are summarized in Table \ref{tab:sum}.

\subsection{Emission from the Gas Disk}
In addition to the X-ray emission from the innermost hot flow, the gas reservoir dominated by the gas disk also offers important observational insights.
The cold gas in the gas disk may be detectable by (sub)millimeter facilities such as ALMA and NOEMA, which trace CO emission.
However, no cold gas measurements or observational constraints are currently available for M60-UCD1.

A fraction of the gas disk can be photoionized by radiation from the old stellar population.
Following the photoionization model used in \citet{2025ApJ...988...68S}, we predict the intrinsic H$\alpha$ and [O\,{\sc iii}]$\lambda5007$ luminosities, prior to dust extinction, to be $1.7\times10^{36}~\rm erg~s^{-1}$ and $1.5\times10^{36}~\rm erg~s^{-1}$, respectively.
The line emission spans a spatial scale of $\sim20~\rm pc$, which is comparable to that of the gas disk and corresponds to an angular size of 0.25 arcsec at the distance of M60-UCD1 \citep[16.5 Mpc;][]{2009ApJ...694..556B}.
Notably, the predicted [O\,{\sc iii}] luminosity is slightly below the [O\,{\sc iii}] detection limit of $\sim4\times10^{36}~\rm erg~s^{-1}$ based on MMT/Hectospec observations on M60-UCD1 \citep{2013ApJ...775L...6S}, as estimated by extrapolating the rough correlation between detection limit and broadband brightness from the MMT/Hectospec survey of globular clusters and UCDs in the Virgo cluster \citep{2019ApJ...885..145S}.
Moreover, unexpectedly strong H30$\alpha$ line emission at 1.3 mm is found amplified by a factor of more than 80 from the cool gas disk in Sgr A* \citep{2019Natur.570...83M,2021ApJ...910..143C}.
This suggests that the H30$\alpha$ line could serve as a potentially powerful probe of the gaseous disk in the wind-fed scenario.
However, the mechanism responsible for the amplification, and whether a similar effect would operate in the case of M60-UCD1, remains uncertain.
Taken together, these considerations call for further analysis and observations to better validate the wind-fed scenario and provide important diagnostics of SMBH accretion in M60-UCD1. 

\section{Discussion}\label{sec:discussion}
\subsection{Comparison with X-ray Observations}\label{sec:discussion:xray}
\begin{figure}[hbpt!]
	\centering
	\includegraphics[width=0.47\textwidth]{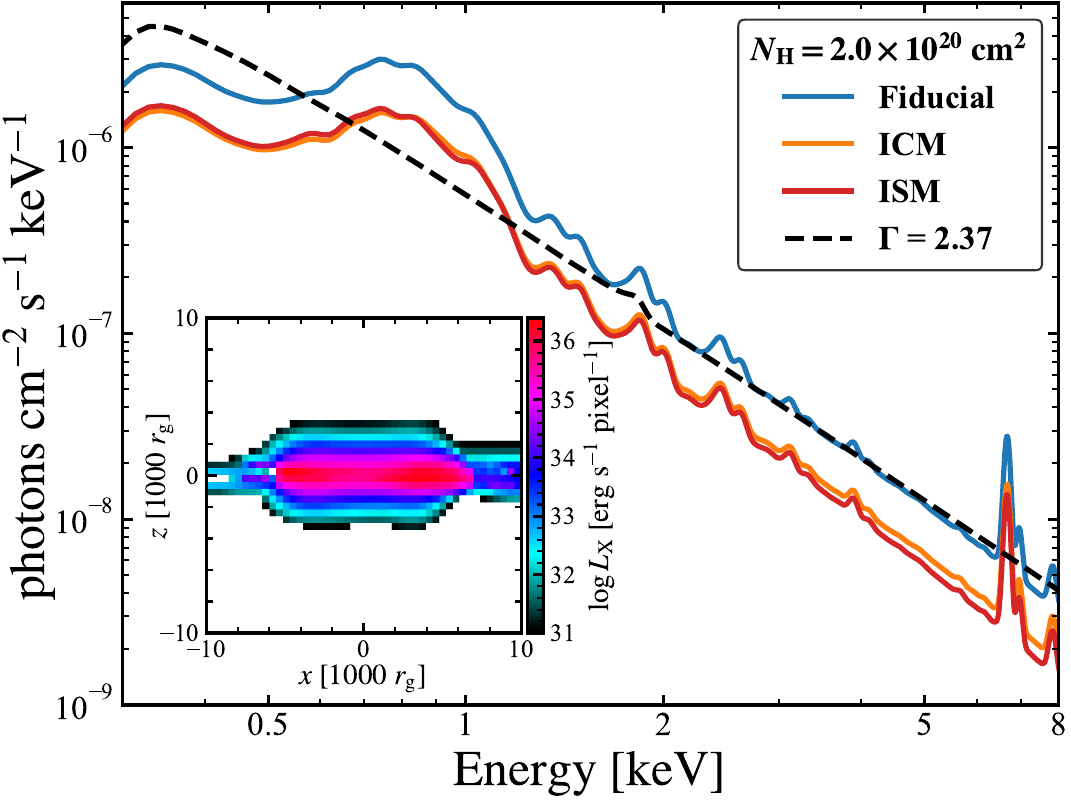}
	\caption{Synthetic X-ray spectra of the last snapshot. The spectra are convolved with the {\it Chandra}/ACIS response matrix with an energy resolution of $\sim100$ eV and a Galactic foreground absorption of $N_{\rm H}=2.0\times10^{20}~\rm cm^2$ suitable for the sky position of M60-UCD1. The black dashed line represents a power-law spectral analog with $\Gamma=2.37$. The inset displays the edge-on X-ray (0.3--8 keV) brightness of the innermost hot corona, with a pixel size of $4\times10^{-4}~{\rm pc}\approx 430~r_{\rm g}$. 
    \label{fig:synthetic_spectra}}
\end{figure}

As shown in Figure \ref{fig:accrate_lx}, the {\tt Fiducial} simulation yields an X-ray luminosity that is in good agreement with the X-ray counterpart of M60-UCD1 ($9.5\times10^{37}~\rm erg~s^{-1}$; \citealt{2013ApJS..204...14L}).
In contrast, the X-ray luminosities of the {\tt ICM/ISM} simulations are slightly lower, by a about factor of two.
The synthetic X-ray spectrum (Figure \ref{fig:synthetic_spectra}) of the {\tt Fiducial} simulation, having a luminosity-weighted temperature of 2.7 keV, can be approximated by a power-law spectrum with a photon-index of 2.37.
The simulated spectra are in broad agreement with the observed spectrum M60-UCD1 of with best-fit photon-index of $\sim1.8$ from the literature \citep{2013ApJ...775L...6S, 2016ApJ...819..164H, 2018ApJ...858..102A}.
We note that this discrepancy in spectral shape can be further mitigated as the spectra are expected to be harder for simulations with higher resolution, which ought to produce gas with higher temperatures and thus harder X-ray photons.
Based on these findings, we conclude that the wind-fed accretion onto the SMBH can explain the observed X-ray emission from M60-UCD1.

The X-ray line-of-sight absorption within the simulation domain is negligible compared to the Galactic foreground absorption for most viewing angles, with a neutral hydrogen column density of $N_{\rm H}\lesssim10^{19}~\rm cm^2$, here effectively including the gas cells with a temperature below $10^4$ K.
However, the thin, cold disk can cause significant absorption, with $N_{\rm H}$ ranging from $2\times10^{23}~\rm cm^{2}$ for an edge-on view to $4\times10^{20}~\rm cm^{2}$ for a viewing angle of $5\arcdeg$.
Therefore, there remains a possibility that the cold disk, despite having a low covering factor of $0.08$, could attenuate the observed X-ray flux from the SMBH accretion by up to two orders of magnitude.

Regarding the ambient environment of M60-UCD1, we find that the {\tt ICM} and {\tt ISM} simulations yield nearly identical X-ray luminosities and spectral properties.
These luminosities are only slightly lower than that of the {\tt Fiducial} simulation and the observed value.
Given the uncertainties stemming from numerical resolution, gas metallicity, and inflowing medium, it is challenging to precisely determine the source of the ram pressure influencing the accretion.

Although our simulations suggest a direct connection between the observed X-ray emission and the SMBH accretion, the LMXB scenario cannot be completely excluded based on current observations and simulations.
Additionally, it is possible that the X-ray emission could result from a combination of the SMBH accretion and stellar sources.
Deep radio observations, ideally simultaneous with X-ray observations, will be crucial in revealing the true nature of the emission.

\subsubsection{Potential Role of Magnetic Fields}\label{subsubsec:bfield}
Additionally, we notice that the X-ray fluctuations in all simulations are significantly weaker than the observed variations of $\sim40\%$ over timescales of months to years (Figure \ref{fig:accrate_lx}).
This might be attributed to the absence of magnetic fields in our simulations. Fluctuations in the accretion rate and X-ray emission could be enhanced through turbulence driven by the magnetorotational instability \citep{1991ApJ...376..214B,1998RvMP...70....1B}, instantaneous energy release via magnetic reconnection \citep[e.g.,][]{2009MNRAS.395.2183Y}, and/or redistributing gas near the BH via magnetic flux eruptions \citep[e.g.,][]{2011MNRAS.418L..79T,2022ApJ...941...30C,2025ApJ...991...89C}.
Furthermore, magnetic fields might play an important role in transporting angular momentum via the magnetorotational instability.
Exploration of the role of magnetic fields is reserved for future work.

\subsection{Role of Ram Pressure Stripping in SMBH Feeding}
There is a long-standing controversy on whether the SMBH activity is enhanced by the ram pressure stripping process.
On the one hand, as a key mechanism for removing gas from galaxies, the ram pressure stripping process would reduce the overall gas supply to the SMBH accretion, thereby suppressing SMBH activity \citep[e.g.,][]{2014MNRAS.437.1942E, 2023A&A...675A..41R}.
On the other hand, numerical simulations indicate that ram pressure stripping process can also reduce the angular momentum of galactic gas, compress the gas, and potentially enhance the accretion onto the SMBH \citep[e.g.,][]{2001MNRAS.328..185S,2009ApJ...694..789T,2018MNRAS.474.3615M,2018MNRAS.476.3781R,2020ApJ...895L...8R,2025MNRAS.542.1901K}.
Observational studies have shown that strongly ram-pressure-stripped galaxies, the so-called jellyfish galaxies, 
exhibit an enhancement in AGN activity and an increased AGN fraction \citep{2017Natur.548..304P,2022ApJ...927..130P}.
Besides the negative and positive effects, there are works suggesting the influence of ram pressure stripping process on SMBH activity is not prominent \citep{2019MNRAS.484..892R, 2025ApJ...979..134T}. 
In the context of UCDs, we note a scarcity of both theoretical and observational studies on the impact of ram pressure stripping, likely due to their compact sizes and low gas content.

In the {\tt ICM/ISM} simulations, the ram pressure stripping process has a slightly suppressive effect on the SMBH accretion and activity (Figure \ref{fig:accrate_lx}).
Compared to the {\tt Fiducial} simulation, gas is removed by the ram pressure instead of being funneled to the SMBH, which is manifested by the lower mass of the gas disk and the lower gas density in the vicinity of the SMBH.
The negative effect may arise due to the low amount of cold gas retained in the simulations of $\sim1000~\rm M_\odot$ that can be funneled to the center.
Additionally, the low stellar mass of M60-UCD1 may limit the impact of ram pressure stripping.
In contrast, the presence of an enhanced AGN fraction is typically observed among massive ram-pressure-stripping galaxies with stellar masses above $\sim10^{10}~\rm M_\odot$ \citep{2017Natur.548..304P, 2022ApJ...927..130P}, and a dependence on galaxy stellar mass has also been suggested \citep[e.g.,][]{2023A&A...671A.118C}.
Similarly, \citet{2025MNRAS.542.1901K} found that the AGN enhancement in jellyfish galaxies in TNG50 is weaker at the lower mass end of $\sim10^{9.5}~\rm M_\odot$.

In addition, the slight decrease in accretion rate and X-ray luminosity by a factor of two in the {\tt ICM/ISM} runs suggests that the environmental effect might be only marginally significant for the wind-fed accretion of SMBHs in UCDs.

\subsection{Implications of Active Galactic Nuclei in UCDs}
Through dedicated hydrodynamical simulations, we have demonstrated that the wind-fed scenario could be promising for the SMBH in M60-UCD1.
Here we discuss the feasibility of this scenario in the more general context of UCDs.
The existence of an SMBH along with its putative activity, i.e., an X-ray/radio counterpart, are determined for only a few UCDs.
Up to now, five Virgo UCDs (M60-UCD1, \citealt{2014Natur.513..398S}; M59-UCD3, \citealt{2018ApJ...858..102A}; M59cO, VUCD3, \citealt{2017ApJ...839...72A}; UCD736, \citealt{2025ApJ...991L..24T}) and one Fornax UCD (Fornax UCD3, \citealt{2018MNRAS.477.4856A}) are found to host a SMBH with $M_{\rm BH}>10^{6}~\rm M_\odot$.
Among these UCDs, M60-UCD1 and M59-UCD3 are coincident with an X-ray source with $L_{\rm X}\sim10^{38}~\rm erg~s^{-1}$ while M59cO is detected in the radio band.
For other UCDs, there are either non-detections or no deep observations available yet.
Besides UCDs with robust dynamical evidences for the presence of SMBHs, NGC5128 UCD320 with non-detection of a central BH ($M_{\rm BH}<10^{6}~\rm M_\odot$) is associated with a flaring X-ray source whose X-ray luminosities and flare timescale can be well explained by an accreting $\lesssim10^6~M_\odot$ BH \citep{2016Natur.538..356I}.
These findings hint that present observations may have captured the signatures of the weakly accreting SMBHs.

Interestingly, M59-UCD3, as an analogue to M60-UCD1 in stellar mass ($1.9\times10^8~\rm M_\odot$), effective radius (27 pc), and black hole mass ($4.2_{-1.7}^{+2.1}\times10^{6}~\rm M_\odot$), has a 0.5--10 keV X-ray luminosity of $1.0\times10^{38}~\rm erg~s^{-1}$ that is also comparable to that of M60-UCD1.
Therefore, M59-UCD3 might represent another case for the wind-fed scenario.
Since M60-UCD1 and M59-UCD3 are among the most massive, densest UCDs hosting the heaviest central BHs, they are expected to have the brightest accretion signatures from wind-fed SMBHs.
Our simulations thus put an upper limit for X-ray luminosities from the wind-fed SMBHs in UCDs, with $L_{\rm X}\lesssim10^{38}~\rm M_\odot$.
Unfortunately, this luminosity cannot let these weakly accreting SMBHs stand out from LMXBs.
Above this upper limit, the brightest X-ray counterparts ($L_{\rm X}\gtrsim10^{38}~\rm M_\odot$) of UCDs would more likely originate from LMXBs (Z. Jin et al. in prep.).

For general UCDs, the X-ray counterparts are usually interpreted as LMXBs.
\citet{2025ApJ...984..132F} found no evidence for an excess of X-ray sources in NGC5128 UCDs possibly hosting central MBHs.
They suggest that these MBHs, with typical inferred masses of $\sim2\times10^{5}~\rm M_\odot$, should have Eddington ratios of $\lesssim2\times10^{-6}$ to interpret the non-detection of the excess.
In fact, the SMBH in M60-UCD1 has an Eddington ratio below this upper limit, which indicates that the wind-fed scenario might be plausible for general UCDs.
Systematic simulations and a larger sample of UCDs with confirmed central BHs are needed to investigate the feeding of these BHs and the origin of the X-ray counterparts.

\section{Conclusions}\label{sec:conclusion}
In this study, we extend the wind-fed mechanism for quiescent SMBHs, where the stellar winds from the NSCs fuel the SMBH, from nearby NSCs such those in the Milky Way and M31, to the case of UCDs.
Some UCDs are well positioned to apply and examine this wind-fed scenario as they are likely the stripped nuclei of dwarf galaxies, host predominantly old stellar populations, lack a significant gas reservoir, and may harbor overmassive BHs.
We perform hydrodynamical simulations on the central SMBH of M60-UCD1 fed by stellar winds from about 1500 AGB stars, and include the putative impact of ram pressure from the interstellar medium or intracluster medium.
Our main results are summarized as follows:

\begin{itemize}
	\item With an accretion radius of $430~\rm r_{\rm g}$ in the simulation, the SMBH with mass of $1.85\times10^{7}~\rm M_\odot$ will have an accretion rate of $\sim10^{-5}~\rm M_\odot~yr^{-1}$ at the quasi-steady state, corresponding to $\sim2\times10^{-5}$ of the Eddington rate. The innermost hot corona ($\sim10^7-10^9~\rm K$) within $10^4~\rm r_{\rm g}$ dominates the X-ray emission with a 0.3--8 keV luminosity of $7\times10^{37}~\rm erg~s^{-1}$ and a photo-index of 2.4 when the synthetic X-ray spectrum is approximated by a power-law. The synthetic X-ray luminosity and spectral shape can naturally explain the X-ray counterpart of M60-UCD1, although contribution by LMXBs cannot be completely ruled out.
	\item Ram pressure from the ICM of Virgo cluster or the ISM of M60 can slightly reduce the SMBH accretion. Both the {\tt ICM} and {\tt ISM} simulations result in a factor of two decrease in the accretion rate and X-ray luminosity, compared to the {\tt Fiducial} simulation without any external inflow.
	\item Within $\sim 5~\rm Myr$, the majority of the wind material settles in a cold gas disk with mass of $1400~M_\odot$, $720~M_\odot$, and $350~M_\odot$ in the {\tt Fiducial}, {\tt ICM}, and {\tt ISM} simulation, respectively. Ram pressure from ISM/ICM significantly removes the gas, and strong ISM ram pressure in particular induces an S-shaped warp in the gas disk.
\end{itemize}

Our simulations demonstrate the suitability of the wind-fed scenario for M60-UCD1 and suggest that the X-ray counterpart of M60-UCD1 could be the accretion signature.
However, systematically investigating the wind-fed accretion of massive black holes within UCDs remains challenging due to the limited number of known UCDs with confirmed SMBHs.
In addition to M60-UCD1, M59-UCD3 may serve as another candidate for wind-fed SMBH accretion, which will be the subject of future simulations.
Recent JWST/NIRSpec+IFU observations have highlighted the remarkable capability of these instruments in searching for central massive black holes in these compact galaxies \citep[e.g.,][]{2025ApJ...991L..24T}.
Combined with deep X-ray and radio observations that reveal potential accretion signatures, these efforts hold great promise in solving the mystery of the X-ray emission associated with UCDs and reaching a consensus on the feeding mechanism of massive black holes.

\begin{acknowledgments}
The authors would like to thank Anil Seth for providing the kinematic map of M60-UCD1 and valuable suggestions, Zihao Jin for assistance with the observational data, and Tao Wang for helpful discussions.
The authors are also grateful to the anonymous referee for critical comments and valuable suggestions.  
This work is supported by the National Natural Science Foundation of
China (grant 12225302) and the Fundamental Research Funds for the Central Universities (grant KG202502). 

\end{acknowledgments}

\software{PLUTO \citep{2007ApJS..170..228M},
          AtomDB \citep{2012ApJ...756..128F},
          CLOUDY \citep{2023RMxAA..59..327C}}




\bibliography{sample701}{}

@ARTICLE{2025ApJ...988...68S,
       author = {{Su}, Zhao and {Li}, Zhiyuan and {Li}, Zongnan},
        title = "{Wind-fed Supermassive Black Hole Accretion by the Nuclear Star Cluster: The Case of M31*}",
      journal = {\apj},
         year = 2025,
        month = jul,
       volume = {988},
       number = {1},
          eid = {68},
        pages = {68},
          doi = {10.3847/1538-4357/ade1d5},
archivePrefix = {arXiv},
       eprint = {2506.04778},
 primaryClass = {astro-ph.GA},
       adsurl = {https://ui.adsabs.harvard.edu/abs/2025ApJ...988...68S}
}

@ARTICLE{2008MNRAS.385..647V,
       author = {{van den Bosch}, R.~C.~E. and {van de Ven}, G. and {Verolme}, E.~K. and {Cappellari}, M. and {de Zeeuw}, P.~T.},
        title = "{Triaxial orbit based galaxy models with an application to the (apparent) decoupled core galaxy NGC 4365}",
      journal = {\mnras},
         year = 2008,
        month = apr,
       volume = {385},
       number = {2},
        pages = {647-666},
          doi = {10.1111/j.1365-2966.2008.12874.x},
archivePrefix = {arXiv},
       eprint = {0712.0113},
 primaryClass = {astro-ph},
       adsurl = {https://ui.adsabs.harvard.edu/abs/2008MNRAS.385..647V}
}

@ARTICLE{2022A&A...667A..51T,
       author = {{Thater}, Sabine and {Jethwa}, Prashin and {Tahmasebzadeh}, Behzad and {Zhu}, Ling and {den Brok}, Mark and {Santucci}, Giulia and {Ding}, Yuchen and {Poci}, Adriano and {Lilley}, Edward and {Tim de Zeeuw}, P. and {Zocchi}, Alice and {Maindl}, Thomas I. and {Rigamonti}, Fabio and {Yang}, Meng and {Fahrion}, Katja and {van de Ven}, Glenn},
        title = "{Testing the robustness of DYNAMITE triaxial Schwarzschild modelling: The effects of correcting the orbit mirroring}",
      journal = {\aap},
         year = 2022,
        month = nov,
       volume = {667},
          eid = {A51},
        pages = {A51},
          doi = {10.1051/0004-6361/202243926},
archivePrefix = {arXiv},
       eprint = {2205.04165},
 primaryClass = {astro-ph.GA},
       adsurl = {https://ui.adsabs.harvard.edu/abs/2022A&A...667A..51T}
}

@software{2020ascl.soft11007J,
       author = {{Jethwa}, Prashin and {Thater}, Sabine and {Maindl}, Thomas and {Van de Ven}, Glenn},
        title = "{DYNAMITE: DYnamics, Age and Metallicity Indicators Tracing Evolution}",
 howpublished = {Astrophysics Source Code Library, record ascl:2011.007},
         year = 2020,
        month = nov,
          eid = {ascl:2011.007},
archivePrefix = {ascl},
       eprint = {2011.007},
       adsurl = {https://ui.adsabs.harvard.edu/abs/2020ascl.soft11007J}
}

@ARTICLE{2014Natur.513..398S,
       author = {{Seth}, Anil C. and {van den Bosch}, Remco and {Mieske}, Steffen and {Baumgardt}, Holger and {Brok}, Mark Den and {Strader}, Jay and {Neumayer}, Nadine and {Chilingarian}, Igor and {Hilker}, Michael and {McDermid}, Richard and {Spitler}, Lee and {Brodie}, Jean and {Frank}, Matthias J. and {Walsh}, Jonelle L.},
        title = "{A supermassive black hole in an ultra-compact dwarf galaxy}",
      journal = {\nat},
         year = 2014,
        month = sep,
       volume = {513},
       number = {7518},
        pages = {398-400},
          doi = {10.1038/nature13762},
archivePrefix = {arXiv},
       eprint = {1409.4769},
 primaryClass = {astro-ph.GA},
       adsurl = {https://ui.adsabs.harvard.edu/abs/2014Natur.513..398S}
}

@ARTICLE{1979ApJ...232..236S,
       author = {{Schwarzschild}, M.},
        title = "{A numerical model for a triaxial stellar system in dynamical equilibrium.}",
      journal = {\apj},
         year = 1979,
        month = aug,
       volume = {232},
        pages = {236-247},
          doi = {10.1086/157282},
       adsurl = {https://ui.adsabs.harvard.edu/abs/1979ApJ...232..236S}
}

@ARTICLE{2007ApJS..170..228M,
       author = {{Mignone}, A. and {Bodo}, G. and {Massaglia}, S. and {Matsakos}, T. and {Tesileanu}, O. and {Zanni}, C. and {Ferrari}, A.},
        title = "{PLUTO: A Numerical Code for Computational Astrophysics}",
      journal = {\apjs},
         year = 2007,
        month = may,
       volume = {170},
       number = {1},
        pages = {228-242},
          doi = {10.1086/513316},
archivePrefix = {arXiv},
       eprint = {astro-ph/0701854},
 primaryClass = {astro-ph},
       adsurl = {https://ui.adsabs.harvard.edu/abs/2007ApJS..170..228M}
}

@ARTICLE{2023RMxAA..59..327C,
       author = {{Chatzikos}, M. and {Bianchi}, S. and {Camilloni}, F. and {Chakraborty}, P. and {Gunasekera}, C.~M. and {Guzm{\'a}n}, F. and {Milby}, J.~S. and {Sarkar}, A. and {Shaw}, G. and {van Hoof}, P.~A.~M. and {Ferland}, G.~J.},
        title = "{The 2023 Release of Cloudy}",
      journal = {\rmxaa},
         year = 2023,
        month = oct,
       volume = {59},
        pages = {327-343},
          doi = {10.22201/ia.01851101p.2023.59.02.12},
archivePrefix = {arXiv},
       eprint = {2308.06396},
 primaryClass = {astro-ph.GA},
       adsurl = {https://ui.adsabs.harvard.edu/abs/2023RMxAA..59..327C}
}

@ARTICLE{2009ApJS..181..391T,
       author = {{Townsend}, R.~H.~D.},
        title = "{An Exact Integration Scheme for Radiative Cooling in Hydrodynamical Simulations}",
      journal = {\apjs},
         year = 2009,
        month = apr,
       volume = {181},
       number = {2},
        pages = {391-397},
          doi = {10.1088/0067-0049/181/2/391},
archivePrefix = {arXiv},
       eprint = {0901.3146},
 primaryClass = {astro-ph.SR},
       adsurl = {https://ui.adsabs.harvard.edu/abs/2009ApJS..181..391T}
}

@ARTICLE{2016ApJS..222....8D,
       author = {{Dotter}, Aaron},
        title = "{MESA Isochrones and Stellar Tracks (MIST) 0: Methods for the Construction of Stellar Isochrones}",
      journal = {\apjs},
         year = 2016,
        month = jan,
       volume = {222},
       number = {1},
          eid = {8},
        pages = {8},
          doi = {10.3847/0067-0049/222/1/8},
archivePrefix = {arXiv},
       eprint = {1601.05144},
 primaryClass = {astro-ph.SR},
       adsurl = {https://ui.adsabs.harvard.edu/abs/2016ApJS..222....8D}
}

@ARTICLE{2016ApJ...823..102C,
       author = {{Choi}, Jieun and {Dotter}, Aaron and {Conroy}, Charlie and {Cantiello}, Matteo and {Paxton}, Bill and {Johnson}, Benjamin D.},
        title = "{Mesa Isochrones and Stellar Tracks (MIST). I. Solar-scaled Models}",
      journal = {\apj},
         year = 2016,
        month = jun,
       volume = {823},
       number = {2},
          eid = {102},
        pages = {102},
          doi = {10.3847/0004-637X/823/2/102},
archivePrefix = {arXiv},
       eprint = {1604.08592},
 primaryClass = {astro-ph.SR},
       adsurl = {https://ui.adsabs.harvard.edu/abs/2016ApJ...823..102C}
}

@ARTICLE{2018ApJ...866...21C,
       author = {{Cummings}, Jeffrey D. and {Kalirai}, Jason S. and {Tremblay}, P. -E. and {Ramirez-Ruiz}, Enrico and {Choi}, Jieun},
        title = "{The White Dwarf Initial-Final Mass Relation for Progenitor Stars from 0.85 to 7.5 M $_{{\ensuremath{\odot}}}$}",
      journal = {\apj},
         year = 2018,
        month = oct,
       volume = {866},
       number = {1},
          eid = {21},
        pages = {21},
          doi = {10.3847/1538-4357/aadfd6},
archivePrefix = {arXiv},
       eprint = {1809.01673},
 primaryClass = {astro-ph.SR},
       adsurl = {https://ui.adsabs.harvard.edu/abs/2018ApJ...866...21C}
}

@ARTICLE{2017ApJ...847...79W,
       author = {{Wood}, R.~A. and {Jones}, C. and {Machacek}, M.~E. and {Forman}, W.~R. and {Bogdan}, A. and {Andrade-Santos}, F. and {Kraft}, R.~P. and {Paggi}, A. and {Roediger}, E.},
        title = "{The Infall of the Virgo Elliptical Galaxy M60 toward M87 and the Gaseous Structures Produced by Kelvin-Helmholtz Instabilities}",
      journal = {\apj},
         year = 2017,
        month = sep,
       volume = {847},
       number = {1},
          eid = {79},
        pages = {79},
          doi = {10.3847/1538-4357/aa8723},
archivePrefix = {arXiv},
       eprint = {1703.05883},
 primaryClass = {astro-ph.GA},
       adsurl = {https://ui.adsabs.harvard.edu/abs/2017ApJ...847...79W}
}

@ARTICLE{2014ApJ...787..134P,
       author = {{Paggi}, Alessandro and {Fabbiano}, Giuseppina and {Kim}, Dong-Woo and {Pellegrini}, Silvia and {Civano}, Francesca and {Strader}, Jay and {Luo}, Bin},
        title = "{Active Galactic Nucleus Feedback in the Hot Halo of NGC 4649}",
      journal = {\apj},
         year = 2014,
        month = jun,
       volume = {787},
       number = {2},
          eid = {134},
        pages = {134},
          doi = {10.1088/0004-637X/787/2/134},
       adsurl = {https://ui.adsabs.harvard.edu/abs/2014ApJ...787..134P}
}

@ARTICLE{2008ApJ...674..869H,
       author = {{Hwang}, Ho Seong and {Lee}, Myung Gyoon and {Park}, Hong Soo and {Kim}, Sang Chul and {Park}, Jang-Hyun and {Sohn}, Young-Jong and {Lee}, Sang-Gak and {Rey}, Soo-Chang and {Lee}, Young-Wook and {Kim}, Ho-Il},
        title = "{The Globular Cluster System of M60 (NGC 4649). II. Kinematics of the Globular Cluster System}",
      journal = {\apj},
         year = 2008,
        month = feb,
       volume = {674},
       number = {2},
        pages = {869-885},
          doi = {10.1086/524001},
archivePrefix = {arXiv},
       eprint = {0709.4309},
 primaryClass = {astro-ph},
       adsurl = {https://ui.adsabs.harvard.edu/abs/2008ApJ...674..869H}
}

@ARTICLE{2024ApJ...974...99B,
       author = {{Balakrishnan}, Mayura and {Russell}, Christopher M.~P. and {Corrales}, Lia and {Calder{\'o}n}, Diego and {Cuadra}, Jorge and {Haggard}, Daryl and {Markoff}, Sera and {Neilsen}, Joey and {Nowak}, Michael and {Wang}, Q. Daniel and {Baganoff}, Frederick},
        title = "{Multistructured Accretion Flow of Sgr A*. II. Signatures of a Cool Accretion Disk in Hydrodynamic Simulations of Stellar Winds}",
      journal = {\apj},
         year = 2024,
        month = oct,
       volume = {974},
       number = {1},
          eid = {99},
        pages = {99},
          doi = {10.3847/1538-4357/ad6866},
archivePrefix = {arXiv},
       eprint = {2406.14631},
 primaryClass = {astro-ph.HE},
       adsurl = {https://ui.adsabs.harvard.edu/abs/2024ApJ...974...99B}
}

@ARTICLE{2025A&A...693A.180C,
       author = {{Calder{\'o}n}, Diego and {Cuadra}, Jorge and {Russell}, Christopher M.~P. and {Burkert}, Andreas and {Rosswog}, Stephan and {Balakrishnan}, Mayura},
        title = "{The formation and stability of a cold disc made out of stellar winds in the Galactic centre}",
      journal = {\aap},
         year = 2025,
        month = jan,
       volume = {693},
          eid = {A180},
        pages = {A180},
          doi = {10.1051/0004-6361/202452800},
archivePrefix = {arXiv},
       eprint = {2411.00100},
 primaryClass = {astro-ph.GA},
       adsurl = {https://ui.adsabs.harvard.edu/abs/2025A&A...693A.180C}
}

@ARTICLE{2014MNRAS.440L..21H,
       author = {{Haan}, S. and {Braun}, R.},
        title = "{On the formation of warped gas discs in galaxies.}",
      journal = {\mnras},
         year = 2014,
        month = may,
       volume = {440},
        pages = {L21-L25},
          doi = {10.1093/mnrasl/slu008},
       adsurl = {https://ui.adsabs.harvard.edu/abs/2014MNRAS.440L..21H}
}

@ARTICLE{1998A&A...337....9R,
       author = {{Reshetnikov}, Vladimir and {Combes}, Francoise},
        title = "{Statistics of optical WARPS in spiral disks}",
      journal = {\aap},
         year = 1998,
        month = sep,
       volume = {337},
        pages = {9-16},
          doi = {10.48550/arXiv.astro-ph/9806114},
archivePrefix = {arXiv},
       eprint = {astro-ph/9806114},
 primaryClass = {astro-ph},
       adsurl = {https://ui.adsabs.harvard.edu/abs/1998A&A...337....9R}
}

@ARTICLE{2009A&A...500..693J,
       author = {{J{\'a}chym}, P. and {K{\"o}ppen}, J. and {Palou{\v{s}}}, J. and {Combes}, F.},
        title = "{Ram pressure stripping of tilted galaxies}",
      journal = {\aap},
         year = 2009,
        month = jun,
       volume = {500},
       number = {2},
        pages = {693-703},
          doi = {10.1051/0004-6361/200811469},
archivePrefix = {arXiv},
       eprint = {0904.3886},
 primaryClass = {astro-ph.CO},
       adsurl = {https://ui.adsabs.harvard.edu/abs/2009A&A...500..693J}
}

@ARTICLE{2006MNRAS.369..567R,
       author = {{Roediger}, Elke and {Br{\"u}ggen}, Marcus},
        title = "{Ram pressure stripping of disc galaxies: the role of the inclination angle}",
      journal = {\mnras},
         year = 2006,
        month = jun,
       volume = {369},
       number = {2},
        pages = {567-580},
          doi = {10.1111/j.1365-2966.2006.10335.x},
archivePrefix = {arXiv},
       eprint = {astro-ph/0512365},
 primaryClass = {astro-ph},
       adsurl = {https://ui.adsabs.harvard.edu/abs/2006MNRAS.369..567R}
}

@ARTICLE{2000Sci...288.1617Q,
       author = {{Quilis}, Vicent and {Moore}, Ben and {Bower}, Richard},
        title = "{Gone with the Wind: The Origin of S0 Galaxies in Clusters}",
      journal = {Science},
         year = 2000,
        month = jun,
       volume = {288},
       number = {5471},
        pages = {1617-1620},
          doi = {10.1126/science.288.5471.1617},
archivePrefix = {arXiv},
       eprint = {astro-ph/0006031},
 primaryClass = {astro-ph},
       adsurl = {https://ui.adsabs.harvard.edu/abs/2000Sci...288.1617Q}
}

@ARTICLE{2025ApJ...979..134T,
       author = {{Tiwari}, Juhi and {Sun}, Ming and {Luo}, Rongxin and {Fossati}, Matteo and {Chen Chien-Ting}, J. and {Tamhane}, Prathamesh},
        title = "{Can Active Galactic Nuclei Activity Be Enhanced by Ram Pressure Stripping?{\textemdash}X-Ray Perspective}",
      journal = {\apj},
         year = 2025,
        month = feb,
       volume = {979},
       number = {2},
          eid = {134},
        pages = {134},
          doi = {10.3847/1538-4357/ad9b7f},
archivePrefix = {arXiv},
       eprint = {2411.05074},
 primaryClass = {astro-ph.GA},
       adsurl = {https://ui.adsabs.harvard.edu/abs/2025ApJ...979..134T}
}

@ARTICLE{2019MNRAS.484..892R,
       author = {{Roman-Oliveira}, Fernanda V. and {Chies-Santos}, Ana L. and {Rodr{\'\i}guez del Pino}, Bruno and {Arag{\'o}n-Salamanca}, A. and {Gray}, Meghan E. and {Bamford}, Steven P.},
        title = "{OMEGA-OSIRIS mapping of emission-line galaxies in A901/2-V. The rich population of jellyfish galaxies in the multicluster system Abell 901/2}",
      journal = {\mnras},
         year = 2019,
        month = mar,
       volume = {484},
       number = {1},
        pages = {892-905},
          doi = {10.1093/mnras/stz007},
archivePrefix = {arXiv},
       eprint = {1812.05629},
 primaryClass = {astro-ph.GA},
       adsurl = {https://ui.adsabs.harvard.edu/abs/2019MNRAS.484..892R}
}

@ARTICLE{2013ApJS..204...14L,
       author = {{Luo}, B. and {Fabbiano}, G. and {Strader}, J. and {Kim}, D. -W. and {Brodie}, J.~P. and {Fragos}, T. and {Gallagher}, J.~S. and {King}, A. and {Zezas}, A.},
        title = "{Deep Chandra Monitoring Observations of NGC 4649. I. Catalog of Source Properties}",
      journal = {\apjs},
         year = 2013,
        month = feb,
       volume = {204},
       number = {2},
          eid = {14},
        pages = {14},
          doi = {10.1088/0067-0049/204/2/14},
archivePrefix = {arXiv},
       eprint = {1211.2804},
 primaryClass = {astro-ph.CO},
       adsurl = {https://ui.adsabs.harvard.edu/abs/2013ApJS..204...14L}
}

@ARTICLE{2013ApJ...775L...6S,
       author = {{Strader}, Jay and {Seth}, Anil C. and {Forbes}, Duncan A. and {Fabbiano}, Giuseppina and {Romanowsky}, Aaron J. and {Brodie}, Jean P. and {Conroy}, Charlie and {Caldwell}, Nelson and {Pota}, Vincenzo and {Usher}, Christopher and {Arnold}, Jacob A.},
        title = "{The Densest Galaxy}",
      journal = {\apjl},
         year = 2013,
        month = sep,
       volume = {775},
       number = {1},
          eid = {L6},
        pages = {L6},
          doi = {10.1088/2041-8205/775/1/L6},
archivePrefix = {arXiv},
       eprint = {1307.7707},
 primaryClass = {astro-ph.CO},
       adsurl = {https://ui.adsabs.harvard.edu/abs/2013ApJ...775L...6S}
}

@ARTICLE{2016ApJ...819..162P,
       author = {{Pandya}, Viraj and {Mulchaey}, John and {Greene}, Jenny E.},
        title = "{A Comprehensive Archival Chandra Search for X-Ray Emission from Ultracompact Dwarf Galaxies}",
      journal = {\apj},
         year = 2016,
        month = mar,
       volume = {819},
       number = {2},
          eid = {162},
        pages = {162},
          doi = {10.3847/0004-637X/819/2/162},
archivePrefix = {arXiv},
       eprint = {1601.01690},
 primaryClass = {astro-ph.GA},
       adsurl = {https://ui.adsabs.harvard.edu/abs/2016ApJ...819..162P}
}

@ARTICLE{2018ApJ...858..102A,
       author = {{Ahn}, Christopher P. and {Seth}, Anil C. and {Cappellari}, Michele and {Krajnovi{\'c}}, Davor and {Strader}, Jay and {Voggel}, Karina T. and {Walsh}, Jonelle L. and {Bahramian}, Arash and {Baumgardt}, Holger and {Brodie}, Jean and {Chilingarian}, Igor and {Chomiuk}, Laura and {den Brok}, Mark and {Frank}, Matthias and {Hilker}, Michael and {McDermid}, Richard M. and {Mieske}, Steffen and {Neumayer}, Nadine and {Nguyen}, Dieu D. and {Pechetti}, Renuka and {Romanowsky}, Aaron J. and {Spitler}, Lee},
        title = "{The Black Hole in the Most Massive Ultracompact Dwarf Galaxy M59-UCD3}",
      journal = {\apj},
         year = 2018,
        month = may,
       volume = {858},
       number = {2},
          eid = {102},
        pages = {102},
          doi = {10.3847/1538-4357/aabc57},
archivePrefix = {arXiv},
       eprint = {1804.02399},
 primaryClass = {astro-ph.GA},
       adsurl = {https://ui.adsabs.harvard.edu/abs/2018ApJ...858..102A}
}

@ARTICLE{2021MNRAS.506.4702F,
       author = {{Ferr{\'e}-Mateu}, A. and {Mezcua}, M. and {Barrows}, R.~S.},
        title = "{A search for active galactic nuclei in low-mass compact galaxies}",
      journal = {\mnras},
         year = 2021,
        month = oct,
       volume = {506},
       number = {4},
        pages = {4702-4714},
          doi = {10.1093/mnras/stab1915},
archivePrefix = {arXiv},
       eprint = {2107.02141},
 primaryClass = {astro-ph.GA},
       adsurl = {https://ui.adsabs.harvard.edu/abs/2021MNRAS.506.4702F}
}

@ARTICLE{2016ApJ...819..164H,
       author = {{Hou}, Meicun and {Li}, Zhiyuan},
        title = "{Chandra Detection of X-Ray Emission from Ultracompact Dwarf Galaxies and Extended Star Clusters}",
      journal = {\apj},
         year = 2016,
        month = mar,
       volume = {819},
       number = {2},
          eid = {164},
        pages = {164},
          doi = {10.3847/0004-637X/819/2/164},
archivePrefix = {arXiv},
       eprint = {1602.01199},
 primaryClass = {astro-ph.GA},
       adsurl = {https://ui.adsabs.harvard.edu/abs/2016ApJ...819..164H}
}

@ARTICLE{2017ApJ...836..104X,
       author = {{Xie}, Fu-Guo and {Yuan}, Feng},
        title = "{Fundamental Plane of Black Hole Activity in the Quiescent Regime}",
      journal = {\apj},
         year = 2017,
        month = feb,
       volume = {836},
       number = {1},
          eid = {104},
        pages = {104},
          doi = {10.3847/1538-4357/aa5b90},
archivePrefix = {arXiv},
       eprint = {1701.06143},
 primaryClass = {astro-ph.HE},
       adsurl = {https://ui.adsabs.harvard.edu/abs/2017ApJ...836..104X}
}

@ARTICLE{2025ApJ...991L..24T,
       author = {{Taylor}, Matthew A. and {Tahmasebzadeh}, Behzad and {Thompson}, Solveig and {Vasiliev}, Eugene and {Valluri}, Monica and {Drinkwater}, Michael J. and {C{\^o}t{\'e}}, Patrick and {Ferrarese}, Laura and {Roediger}, Joel and {Baumgardt}, Holger and {Bentz}, Misty C. and {Dage}, Kristen and {Peng}, Eric W. and {Lapeer}, Drew and {Liu}, Chengze and {Sumners}, Zach and {Wang}, Kaixiang and {Baldassare}, Vivienne and {Blakeslee}, John P. and {Ko}, Youkyung and {Woods}, Tyrone E.},
        title = "{A Supermassive Black Hole in a Diminutive Ultracompact Dwarf Galaxy Discovered with JWST/NIRSpec+IFU}",
      journal = {\apjl},
         year = 2025,
        month = sep,
       volume = {991},
       number = {1},
          eid = {L24},
        pages = {L24},
          doi = {10.3847/2041-8213/ae028e},
archivePrefix = {arXiv},
       eprint = {2503.00113},
 primaryClass = {astro-ph.GA},
       adsurl = {https://ui.adsabs.harvard.edu/abs/2025ApJ...991L..24T}
}

@ARTICLE{2017ApJ...839...72A,
       author = {{Ahn}, Christopher P. and {Seth}, Anil C. and {den Brok}, Mark and {Strader}, Jay and {Baumgardt}, Holger and {van den Bosch}, Remco and {Chilingarian}, Igor and {Frank}, Matthias and {Hilker}, Michael and {McDermid}, Richard and {Mieske}, Steffen and {Romanowsky}, Aaron J. and {Spitler}, Lee and {Brodie}, Jean and {Neumayer}, Nadine and {Walsh}, Jonelle L.},
        title = "{Detection of Supermassive Black Holes in Two Virgo Ultracompact Dwarf Galaxies}",
      journal = {\apj},
         year = 2017,
        month = apr,
       volume = {839},
       number = {2},
          eid = {72},
        pages = {72},
          doi = {10.3847/1538-4357/aa6972},
archivePrefix = {arXiv},
       eprint = {1703.09221},
 primaryClass = {astro-ph.GA},
       adsurl = {https://ui.adsabs.harvard.edu/abs/2017ApJ...839...72A}
}

@ARTICLE{2018MNRAS.477.4856A,
       author = {{Afanasiev}, Anton V. and {Chilingarian}, Igor V. and {Mieske}, Steffen and {Voggel}, Karina T. and {Picotti}, Arianna and {Hilker}, Michael and {Seth}, Anil and {Neumayer}, Nadine and {Frank}, Matthias and {Romanowsky}, Aaron J. and {Hau}, George and {Baumgardt}, Holger and {Ahn}, Christopher and {Strader}, Jay and {den Brok}, Mark and {McDermid}, Richard and {Spitler}, Lee and {Brodie}, Jean and {Walsh}, Jonelle L.},
        title = "{A 3.5 million Solar masses black hole in the centre of the ultracompact dwarf galaxy fornax UCD3}",
      journal = {\mnras},
         year = 2018,
        month = jul,
       volume = {477},
       number = {4},
        pages = {4856-4865},
          doi = {10.1093/mnras/sty913},
archivePrefix = {arXiv},
       eprint = {1804.02938},
 primaryClass = {astro-ph.GA},
       adsurl = {https://ui.adsabs.harvard.edu/abs/2018MNRAS.477.4856A}
}

@ARTICLE{2016Natur.538..356I,
       author = {{Irwin}, Jimmy A. and {Maksym}, W. Peter and {Sivakoff}, Gregory R. and {Romanowsky}, Aaron J. and {Lin}, Dacheng and {Speegle}, Tyler and {Prado}, Ian and {Mildebrath}, David and {Strader}, Jay and {Liu}, Jifeng and {Miller}, Jon M.},
        title = "{Ultraluminous X-ray bursts in two ultracompact companions to nearby elliptical galaxies}",
      journal = {\nat},
         year = 2016,
        month = oct,
       volume = {538},
       number = {7625},
        pages = {356-358},
          doi = {10.1038/nature19822},
archivePrefix = {arXiv},
       eprint = {1610.05781},
 primaryClass = {astro-ph.HE},
       adsurl = {https://ui.adsabs.harvard.edu/abs/2016Natur.538..356I}
}

@ARTICLE{2012ApJ...747...72D,
       author = {{Dabringhausen}, J{\"o}rg and {Kroupa}, Pavel and {Pflamm-Altenburg}, Jan and {Mieske}, Steffen},
        title = "{Low-mass X-Ray Binaries Indicate a Top-heavy Stellar Initial Mass Function in Ultracompact Dwarf Galaxies}",
      journal = {\apj},
         year = 2012,
        month = mar,
       volume = {747},
       number = {1},
          eid = {72},
        pages = {72},
          doi = {10.1088/0004-637X/747/1/72},
archivePrefix = {arXiv},
       eprint = {1110.2779},
 primaryClass = {astro-ph.CO},
       adsurl = {https://ui.adsabs.harvard.edu/abs/2012ApJ...747...72D}
}

@ARTICLE{2025ApJ...984..132F,
       author = {{Feyan}, Samuel L. and {Urquhart}, Ryan and {Strader}, Jay and {Seth}, Anil C. and {Sand}, David J. and {Caldwell}, Nelson and {Crnojevi{\'c}}, Denija and {Dumont}, Antoine and {Voggel}, Karina},
        title = "{X-Ray Constraints on Wandering Black Holes in Stripped Galaxy Nuclei in the Halo of NGC 5128}",
      journal = {\apj},
         year = 2025,
        month = may,
       volume = {984},
       number = {2},
          eid = {132},
        pages = {132},
          doi = {10.3847/1538-4357/adc7b9},
archivePrefix = {arXiv},
       eprint = {2504.08034},
 primaryClass = {astro-ph.GA},
       adsurl = {https://ui.adsabs.harvard.edu/abs/2025ApJ...984..132F}
}

@ARTICLE{2017Natur.548..304P,
       author = {{Poggianti}, Bianca M. and {Jaff{\'e}}, Yara L. and {Moretti}, Alessia and {Gullieuszik}, Marco and {Radovich}, Mario and {Tonnesen}, Stephanie and {Fritz}, Jacopo and {Bettoni}, Daniela and {Vulcani}, Benedetta and {Fasano}, Giovanni and {Bellhouse}, Callum and {Hau}, George and {Omizzolo}, Alessandro},
        title = "{Ram-pressure feeding of supermassive black holes}",
      journal = {\nat},
         year = 2017,
        month = aug,
       volume = {548},
       number = {7667},
        pages = {304-309},
          doi = {10.1038/nature23462},
archivePrefix = {arXiv},
       eprint = {1708.09036},
 primaryClass = {astro-ph.GA},
       adsurl = {https://ui.adsabs.harvard.edu/abs/2017Natur.548..304P}
}

@ARTICLE{2022ApJ...927..130P,
       author = {{Peluso}, Giorgia and {Vulcani}, Benedetta and {Poggianti}, Bianca M. and {Moretti}, Alessia and {Radovich}, Mario and {Smith}, Rory and {Jaff{\'e}}, Yara L. and {Crossett}, Jacob and {Gullieuszik}, Marco and {Fritz}, Jacopo and {Ignesti}, Alessandro},
        title = "{Exploring the AGN-Ram Pressure Stripping Connection in Local Clusters}",
      journal = {\apj},
         year = 2022,
        month = mar,
       volume = {927},
       number = {1},
          eid = {130},
        pages = {130},
          doi = {10.3847/1538-4357/ac4225},
archivePrefix = {arXiv},
       eprint = {2111.02538},
 primaryClass = {astro-ph.GA},
       adsurl = {https://ui.adsabs.harvard.edu/abs/2022ApJ...927..130P}
}

@ARTICLE{2025MNRAS.542.1901K,
       author = {{Kurinchi-Vendhan}, Shalini and {Rohr}, Eric and {Pillepich}, Annalisa and {Zinger}, Elad and {Ayromlou}, Mohammadreza and {Joshi}, Gandhali D.},
        title = "{Jellyfish galaxies with the IllustrisTNG simulations {\textendash} Supermassive black hole activity in dense environments with ram-pressure stripped satellites}",
      journal = {\mnras},
         year = 2025,
        month = sep,
       volume = {542},
       number = {3},
        pages = {1901-1922},
          doi = {10.1093/mnras/staf1280},
archivePrefix = {arXiv},
       eprint = {2506.05474},
 primaryClass = {astro-ph.GA},
       adsurl = {https://ui.adsabs.harvard.edu/abs/2025MNRAS.542.1901K}
}

@ARTICLE{2023A&A...671A.118C,
       author = {{Cattorini}, Federico and {Gavazzi}, Giuseppe and {Boselli}, Alessandro and {Fossati}, Matteo},
        title = "{A complete spectroscopic catalogue of local galaxies in the northern spring sky: Gas properties and nuclear activity in different environments}",
      journal = {\aap},
         year = 2023,
        month = mar,
       volume = {671},
          eid = {A118},
        pages = {A118},
          doi = {10.1051/0004-6361/202244738},
archivePrefix = {arXiv},
       eprint = {2211.06437},
 primaryClass = {astro-ph.GA},
       adsurl = {https://ui.adsabs.harvard.edu/abs/2023A&A...671A.118C}
}

@ARTICLE{2008ARA&A..46..475H,
       author = {{Ho}, L.~C.},
        title = "{Nuclear activity in nearby galaxies.}",
      journal = {\araa},
         year = 2008,
        month = sep,
       volume = {46},
        pages = {475-539},
          doi = {10.1146/annurev.astro.45.051806.110546},
archivePrefix = {arXiv},
       eprint = {0803.2268},
 primaryClass = {astro-ph},
       adsurl = {https://ui.adsabs.harvard.edu/abs/2008ARA&A..46..475H}
}

@ARTICLE{2019ApJ...885...16Y,
       author = {{Yoon}, Doosoo and {Yuan}, Feng and {Ostriker}, Jeremiah P. and {Ciotti}, Luca and {Zhu}, Bocheng},
        title = "{On the Role of the Hot Feedback Mode in Active Galactic Nuclei Feedback in an Elliptical Galaxy}",
      journal = {\apj},
         year = 2019,
        month = nov,
       volume = {885},
       number = {1},
          eid = {16},
        pages = {16},
          doi = {10.3847/1538-4357/ab45e8},
archivePrefix = {arXiv},
       eprint = {1901.07570},
 primaryClass = {astro-ph.HE},
       adsurl = {https://ui.adsabs.harvard.edu/abs/2019ApJ...885...16Y}
}

@ARTICLE{2017MNRAS.465.3291W,
       author = {{Weinberger}, Rainer and {Springel}, Volker and {Hernquist}, Lars and {Pillepich}, Annalisa and {Marinacci}, Federico and {Pakmor}, R{\"u}diger and {Nelson}, Dylan and {Genel}, Shy and {Vogelsberger}, Mark and {Naiman}, Jill and {Torrey}, Paul},
        title = "{Simulating galaxy formation with black hole driven thermal and kinetic feedback}",
      journal = {\mnras},
         year = 2017,
        month = mar,
       volume = {465},
       number = {3},
        pages = {3291-3308},
          doi = {10.1093/mnras/stw2944},
archivePrefix = {arXiv},
       eprint = {1607.03486},
 primaryClass = {astro-ph.GA},
       adsurl = {https://ui.adsabs.harvard.edu/abs/2017MNRAS.465.3291W}
}

@ARTICLE{2018MNRAS.479.4056W,
       author = {{Weinberger}, Rainer and {Springel}, Volker and {Pakmor}, R{\"u}diger and {Nelson}, Dylan and {Genel}, Shy and {Pillepich}, Annalisa and {Vogelsberger}, Mark and {Marinacci}, Federico and {Naiman}, Jill and {Torrey}, Paul and {Hernquist}, Lars},
        title = "{Supermassive black holes and their feedback effects in the IllustrisTNG simulation}",
      journal = {\mnras},
         year = 2018,
        month = sep,
       volume = {479},
       number = {3},
        pages = {4056-4072},
          doi = {10.1093/mnras/sty1733},
archivePrefix = {arXiv},
       eprint = {1710.04659},
 primaryClass = {astro-ph.GA},
       adsurl = {https://ui.adsabs.harvard.edu/abs/2018MNRAS.479.4056W}
}

@ARTICLE{2020A&ARv..28....4N,
       author = {{Neumayer}, Nadine and {Seth}, Anil and {B{\"o}ker}, Torsten},
        title = "{Nuclear star clusters}",
      journal = {\aapr},
         year = 2020,
        month = jul,
       volume = {28},
       number = {1},
          eid = {4},
        pages = {4},
          doi = {10.1007/s00159-020-00125-0},
archivePrefix = {arXiv},
       eprint = {2001.03626},
 primaryClass = {astro-ph.GA},
       adsurl = {https://ui.adsabs.harvard.edu/abs/2020A&ARv..28....4N}
}

@ARTICLE{2023Natur.623..296W,
       author = {{Wang}, Kaixiang and {Peng}, Eric W. and {Liu}, Chengze and {Mihos}, J. Christopher and {C{\^o}t{\'e}}, Patrick and {Ferrarese}, Laura and {Taylor}, Matthew A. and {Blakeslee}, John P. and {Cuillandre}, Jean-Charles and {Duc}, Pierre-Alain and {Guhathakurta}, Puragra and {Gwyn}, Stephen and {Ko}, Youkyung and {Lan{\c{c}}on}, Ariane and {Lim}, Sungsoon and {MacArthur}, Lauren A. and {Puzia}, Thomas and {Roediger}, Joel and {Sales}, Laura V. and {S{\'a}nchez-Janssen}, Rub{\'e}n and {Spengler}, Chelsea and {Toloba}, Elisa and {Zhang}, Hongxin and {Zhu}, Mingcheng},
        title = "{An evolutionary continuum from nucleated dwarf galaxies to star clusters}",
      journal = {\nat},
         year = 2023,
        month = nov,
       volume = {623},
       number = {7986},
        pages = {296-300},
          doi = {10.1038/s41586-023-06650-z},
archivePrefix = {arXiv},
       eprint = {2311.05448},
 primaryClass = {astro-ph.GA},
       adsurl = {https://ui.adsabs.harvard.edu/abs/2023Natur.623..296W}
}

@ARTICLE{2023MNRAS.526L.136P,
       author = {{Paudel}, Sanjaya and {Duc}, Pierre-Alain and {Lim}, Sungsoon and {Poulain}, M{\'e}lina and {Marleau}, Francine R. and {M{\"u}ller}, Oliver and {S{\'a}nchez-Janssen}, Rub{\'e}n and {Habas}, Rebecca and {Durrell}, Patrick R. and {Heesters}, Nick and {Chhatkuli}, Daya Nidhi and {Yoon}, Suk-Jin},
        title = "{The creation of a massive UCD by tidal threshing from NGC 936}",
      journal = {\mnras},
         year = 2023,
        month = nov,
       volume = {526},
       number = {1},
        pages = {L136-L142},
          doi = {10.1093/mnrasl/slad126},
archivePrefix = {arXiv},
       eprint = {2309.08098},
 primaryClass = {astro-ph.GA},
       adsurl = {https://ui.adsabs.harvard.edu/abs/2023MNRAS.526L.136P}
}

@ARTICLE{2012MNRAS.425..325F,
       author = {{Francis}, K.~J. and {Drinkwater}, M.~J. and {Chilingarian}, Igor V. and {Bolt}, A.~M. and {Firth}, P.},
        title = "{The chemical composition of ultracompact dwarf galaxies in the Virgo and Fornax clusters}",
      journal = {\mnras},
         year = 2012,
        month = sep,
       volume = {425},
       number = {1},
        pages = {325-337},
          doi = {10.1111/j.1365-2966.2012.21465.x},
archivePrefix = {arXiv},
       eprint = {1207.3382},
 primaryClass = {astro-ph.CO},
       adsurl = {https://ui.adsabs.harvard.edu/abs/2012MNRAS.425..325F}
}

@ARTICLE{2015ApJ...812...34L,
       author = {{Liu}, Chengze and {Peng}, Eric W. and {C{\^o}t{\'e}}, Patrick and {Ferrarese}, Laura and {Jord{\'a}n}, Andr{\'e}s and {Mihos}, J. Christopher and {Zhang}, Hong-Xin and {Mu{\~n}oz}, Roberto P. and {Puzia}, Thomas H. and {Lan{\c{c}}on}, Ariane and {Gwyn}, Stephen and {Cuillandre}, Jean-Charles and {Blakeslee}, John P. and {Boselli}, Alessandro and {Durrell}, Patrick R. and {Duc}, Pierre-Alain and {Guhathakurta}, Puragra and {MacArthur}, Lauren A. and {Mei}, Simona and {S{\'a}nchez-Janssen}, Rub{\'e}n and {Xu}, Haiguang},
        title = "{The Next Generation Virgo Cluster Survey. X. Properties of Ultra-compact Dwarfs in the M87, M49, and M60 Regions.}",
      journal = {\apj},
         year = 2015,
        month = oct,
       volume = {812},
       number = {1},
          eid = {34},
        pages = {34},
          doi = {10.1088/0004-637X/812/1/34},
archivePrefix = {arXiv},
       eprint = {1508.07334},
 primaryClass = {astro-ph.GA},
       adsurl = {https://ui.adsabs.harvard.edu/abs/2015ApJ...812...34L}
}

@ARTICLE{2020ApJS..250...17L,
       author = {{Liu}, Chengze and {C{\^o}t{\'e}}, Patrick and {Peng}, Eric W. and {Roediger}, Joel and {Zhang}, Hongxin and {Ferrarese}, Laura and {S{\'a}nchez-Janssen}, Ruben and {Guhathakurta}, Puragra and {Yang}, Xiaohu and {Jing}, Yipeng and {Alamo-Mart{\'\i}nez}, Karla and {Blakeslee}, John P. and {Boselli}, Alessandro and {Cuilandre}, Jean-Charles and {Duc}, Pierre-Alain and {Durrell}, Patrick and {Gwyn}, Stephen and {Jord{\'a}n}, Andres and {Ko}, Youkyung and {Lan{\c{c}}on}, Ariane and {Lim}, Sungsoon and {Longobardi}, Alessia and {Mei}, Simona and {Mihos}, J. Christopher and {Mu{\~n}oz}, Roberto and {Powalka}, Mathieu and {Puzia}, Thomas and {Spengler}, Chelsea and {Toloba}, Elisa},
        title = "{The Next Generation Virgo Cluster Survey. XXXIV. Ultracompact Dwarf Galaxies in the Virgo Cluster}",
      journal = {\apjs},
         year = 2020,
        month = sep,
       volume = {250},
       number = {1},
          eid = {17},
        pages = {17},
          doi = {10.3847/1538-4365/abad91},
archivePrefix = {arXiv},
       eprint = {2007.15275},
 primaryClass = {astro-ph.GA},
       adsurl = {https://ui.adsabs.harvard.edu/abs/2020ApJS..250...17L}
}

@ARTICLE{2009ApJ...699..626H,
       author = {{Ho}, Luis C.},
        title = "{Radiatively Inefficient Accretion in Nearby Galaxies}",
      journal = {\apj},
         year = 2009,
        month = jul,
       volume = {699},
       number = {1},
        pages = {626-637},
          doi = {10.1088/0004-637X/699/1/626},
archivePrefix = {arXiv},
       eprint = {0906.4104},
 primaryClass = {astro-ph.GA},
       adsurl = {https://ui.adsabs.harvard.edu/abs/2009ApJ...699..626H}
}

@ARTICLE{2018MNRAS.478.3544R,
       author = {{Ressler}, S.~M. and {Quataert}, E. and {Stone}, J.~M.},
        title = "{Hydrodynamic simulations of the inner accretion flow of Sagittarius A* fuelled by stellar winds}",
      journal = {\mnras},
         year = 2018,
        month = aug,
       volume = {478},
       number = {3},
        pages = {3544-3563},
          doi = {10.1093/mnras/sty1146},
archivePrefix = {arXiv},
       eprint = {1805.00474},
 primaryClass = {astro-ph.HE},
       adsurl = {https://ui.adsabs.harvard.edu/abs/2018MNRAS.478.3544R}
}

@ARTICLE{2020MNRAS.492.3272R,
       author = {{Ressler}, S.~M. and {Quataert}, E. and {Stone}, J.~M.},
        title = "{The surprisingly small impact of magnetic fields on the inner accretion flow of Sagittarius A* fueled by stellar winds}",
      journal = {\mnras},
         year = 2020,
        month = mar,
       volume = {492},
       number = {3},
        pages = {3272-3293},
          doi = {10.1093/mnras/stz3605},
archivePrefix = {arXiv},
       eprint = {2001.04469},
 primaryClass = {astro-ph.HE},
       adsurl = {https://ui.adsabs.harvard.edu/abs/2020MNRAS.492.3272R}
}

@ARTICLE{2020ApJ...896L...6R,
       author = {{Ressler}, Sean M. and {White}, Christopher J. and {Quataert}, Eliot and {Stone}, James M.},
        title = "{Ab Initio Horizon-scale Simulations of Magnetically Arrested Accretion in Sagittarius A* Fed by Stellar Winds}",
      journal = {\apjl},
         year = 2020,
        month = jun,
       volume = {896},
       number = {1},
          eid = {L6},
        pages = {L6},
          doi = {10.3847/2041-8213/ab9532},
archivePrefix = {arXiv},
       eprint = {2006.00005},
 primaryClass = {astro-ph.HE},
       adsurl = {https://ui.adsabs.harvard.edu/abs/2020ApJ...896L...6R}
}

@ARTICLE{1999A&AS..134...75H,
       author = {{Hilker}, M. and {Infante}, L. and {Vieira}, G. and {Kissler-Patig}, M. and {Richtler}, T.},
        title = "{The central region of the Fornax cluster. II. Spectroscopy and radial velocities of member and background galaxies}",
      journal = {\aaps},
         year = 1999,
        month = jan,
       volume = {134},
        pages = {75-86},
          doi = {10.1051/aas:1999434},
archivePrefix = {arXiv},
       eprint = {astro-ph/9807144},
 primaryClass = {astro-ph},
       adsurl = {https://ui.adsabs.harvard.edu/abs/1999A&AS..134...75H}
}

@ARTICLE{2000PASA...17..227D,
       author = {{Drinkwater}, M.~J. and {Jones}, J.~B. and {Gregg}, M.~D. and {Phillipps}, S.},
        title = "{Compact Stellar Systems in the Fornax Cluster: Super-massive Star Clusters or Extremely Compact Dwarf Galaxies?}",
      journal = {\pasa},
         year = 2000,
        month = dec,
       volume = {17},
       number = {3},
        pages = {227-233},
          doi = {10.1071/AS00034},
archivePrefix = {arXiv},
       eprint = {astro-ph/0002003},
 primaryClass = {astro-ph},
       adsurl = {https://ui.adsabs.harvard.edu/abs/2000PASA...17..227D}
}

@ARTICLE{2001ApJ...560..201P,
       author = {{Phillipps}, S. and {Drinkwater}, M.~J. and {Gregg}, M.~D. and {Jones}, J.~B.},
        title = "{Ultracompact Dwarf Galaxies in the Fornax Cluster}",
      journal = {\apj},
         year = 2001,
        month = oct,
       volume = {560},
       number = {1},
        pages = {201-206},
          doi = {10.1086/322517},
archivePrefix = {arXiv},
       eprint = {astro-ph/0106377},
 primaryClass = {astro-ph},
       adsurl = {https://ui.adsabs.harvard.edu/abs/2001ApJ...560..201P}
}

@ARTICLE{2003Natur.423..519D,
       author = {{Drinkwater}, M.~J. and {Gregg}, M.~D. and {Hilker}, M. and {Bekki}, K. and {Couch}, W.~J. and {Ferguson}, H.~C. and {Jones}, J.~B. and {Phillipps}, S.},
        title = "{A class of compact dwarf galaxies from disruptive processes in galaxy clusters}",
      journal = {\nat},
         year = 2003,
        month = may,
       volume = {423},
       number = {6939},
        pages = {519-521},
          doi = {10.1038/nature01666},
archivePrefix = {arXiv},
       eprint = {astro-ph/0306026},
 primaryClass = {astro-ph},
       adsurl = {https://ui.adsabs.harvard.edu/abs/2003Natur.423..519D}
}

@ARTICLE{2001ApJ...552L.105B,
       author = {{Bekki}, Kenji and {Couch}, Warrick J. and {Drinkwater}, Michael J.},
        title = "{Galaxy Threshing and the Formation of Ultracompact Dwarf Galaxies}",
      journal = {\apjl},
         year = 2001,
        month = may,
       volume = {552},
       number = {2},
        pages = {L105-L108},
          doi = {10.1086/320339},
archivePrefix = {arXiv},
       eprint = {astro-ph/0106402},
 primaryClass = {astro-ph},
       adsurl = {https://ui.adsabs.harvard.edu/abs/2001ApJ...552L.105B}
}

@ARTICLE{2013MNRAS.433.1997P,
       author = {{Pfeffer}, J. and {Baumgardt}, H.},
        title = "{Ultra-compact dwarf galaxy formation by tidal stripping of nucleated dwarf galaxies}",
      journal = {\mnras},
         year = 2013,
        month = aug,
       volume = {433},
       number = {3},
        pages = {1997-2005},
          doi = {10.1093/mnras/stt867},
archivePrefix = {arXiv},
       eprint = {1305.3656},
 primaryClass = {astro-ph.GA},
       adsurl = {https://ui.adsabs.harvard.edu/abs/2013MNRAS.433.1997P}
}

@ARTICLE{2003MNRAS.344..399B,
       author = {{Bekki}, K. and {Couch}, W.~J. and {Drinkwater}, M.~J. and {Shioya}, Y.},
        title = "{Galaxy threshing and the origin of ultra-compact dwarf galaxies in the Fornax cluster}",
      journal = {\mnras},
         year = 2003,
        month = sep,
       volume = {344},
       number = {2},
        pages = {399-411},
          doi = {10.1046/j.1365-8711.2003.06916.x},
archivePrefix = {arXiv},
       eprint = {astro-ph/0308243},
 primaryClass = {astro-ph},
       adsurl = {https://ui.adsabs.harvard.edu/abs/2003MNRAS.344..399B}
}

@ARTICLE{2012ApJ...756..128F,
       author = {{Foster}, A.~R. and {Ji}, L. and {Smith}, R.~K. and {Brickhouse}, N.~S.},
        title = "{Updated Atomic Data and Calculations for X-Ray Spectroscopy}",
      journal = {\apj},
         year = 2012,
        month = sep,
       volume = {756},
       number = {2},
          eid = {128},
        pages = {128},
          doi = {10.1088/0004-637X/756/2/128},
archivePrefix = {arXiv},
       eprint = {1207.0576},
 primaryClass = {astro-ph.HE},
       adsurl = {https://ui.adsabs.harvard.edu/abs/2012ApJ...756..128F}
}

@ARTICLE{2001MNRAS.322..231K,
       author = {{Kroupa}, Pavel},
        title = "{On the variation of the initial mass function}",
      journal = {\mnras},
         year = 2001,
        month = apr,
       volume = {322},
       number = {2},
        pages = {231-246},
          doi = {10.1046/j.1365-8711.2001.04022.x},
archivePrefix = {arXiv},
       eprint = {astro-ph/0009005},
 primaryClass = {astro-ph},
       adsurl = {https://ui.adsabs.harvard.edu/abs/2001MNRAS.322..231K}
}

@ARTICLE{2002Sci...295...82K,
       author = {{Kroupa}, Pavel},
        title = "{The Initial Mass Function of Stars: Evidence for Uniformity in Variable Systems}",
      journal = {Science},
         year = 2002,
        month = jan,
       volume = {295},
       number = {5552},
        pages = {82-91},
          doi = {10.1126/science.1067524},
archivePrefix = {arXiv},
       eprint = {astro-ph/0201098},
 primaryClass = {astro-ph},
       adsurl = {https://ui.adsabs.harvard.edu/abs/2002Sci...295...82K}
}

@ARTICLE{2019ApJ...885..145S,
       author = {{Sun}, Weijia and {Peng}, Eric W. and {Ko}, Youkyung and {C{\^o}t{\'e}}, Patrick and {Ferrarese}, Laura and {Lee}, Myung Gyoon and {Liu}, Chengze and {Longobardi}, Alessia and {Chilingarian}, Igor V. and {Spengler}, Chelsea and {Zabludoff}, Ann I. and {Zhang}, Hong-Xin and {Cuillandre}, Jean-Charles and {Gwyn}, Stephen D.~J.},
        title = "{The Next Generation Virgo Cluster Survey. XVII. A Search for Planetary Nebulae in Virgo Cluster Globular Clusters}",
      journal = {\apj},
         year = 2019,
        month = nov,
       volume = {885},
       number = {2},
          eid = {145},
        pages = {145},
          doi = {10.3847/1538-4357/ab49fb},
archivePrefix = {arXiv},
       eprint = {1910.00169},
 primaryClass = {astro-ph.GA},
       adsurl = {https://ui.adsabs.harvard.edu/abs/2019ApJ...885..145S}
}

@ARTICLE{2023A&A...675A..41R,
       author = {{Rodr{\'\i}guez Del Pino}, B. and {Arribas}, S. and {Chies-Santos}, A.~L. and {Lamperti}, I. and {Perna}, M. and {V{\'\i}lchez}, J.~M.},
        title = "{The impact of environmental effects on active galactic nuclei: A decline in the incidence of ionized outflows}",
      journal = {\aap},
         year = 2023,
        month = jul,
       volume = {675},
          eid = {A41},
        pages = {A41},
          doi = {10.1051/0004-6361/202346051},
archivePrefix = {arXiv},
       eprint = {2304.04786},
 primaryClass = {astro-ph.GA},
       adsurl = {https://ui.adsabs.harvard.edu/abs/2023A&A...675A..41R}
}

@ARTICLE{2014MNRAS.437.1942E,
       author = {{Ehlert}, S. and {von der Linden}, A. and {Allen}, S.~W. and {Brandt}, W.~N. and {Xue}, Y.~Q. and {Luo}, B. and {Mantz}, A. and {Morris}, R.~G. and {Applegate}, D. and {Kelly}, P.},
        title = "{X-ray bright active galactic nuclei in massive galaxy clusters - II. The fraction of galaxies hosting active nuclei}",
      journal = {\mnras},
         year = 2014,
        month = jan,
       volume = {437},
       number = {2},
        pages = {1942-1949},
          doi = {10.1093/mnras/stt2025},
archivePrefix = {arXiv},
       eprint = {1310.5711},
 primaryClass = {astro-ph.CO},
       adsurl = {https://ui.adsabs.harvard.edu/abs/2014MNRAS.437.1942E}
}

@ARTICLE{2020ApJ...895L...8R,
       author = {{Ricarte}, Angelo and {Tremmel}, Michael and {Natarajan}, Priyamvada and {Quinn}, Thomas},
        title = "{A Link between Ram Pressure Stripping and Active Galactic Nuclei}",
      journal = {\apjl},
         year = 2020,
        month = may,
       volume = {895},
       number = {1},
          eid = {L8},
        pages = {L8},
          doi = {10.3847/2041-8213/ab9022},
archivePrefix = {arXiv},
       eprint = {2003.05950},
 primaryClass = {astro-ph.GA},
       adsurl = {https://ui.adsabs.harvard.edu/abs/2020ApJ...895L...8R}
}

@ARTICLE{2018MNRAS.474.3615M,
       author = {{Marshall}, Madeline A. and {Shabala}, Stanislav S. and {Krause}, Martin G.~H. and {Pimbblet}, Kevin A. and {Croton}, Darren J. and {Owers}, Matt S.},
        title = "{Triggering active galactic nuclei in galaxy clusters}",
      journal = {\mnras},
         year = 2018,
        month = mar,
       volume = {474},
       number = {3},
        pages = {3615-3628},
          doi = {10.1093/mnras/stx2996},
archivePrefix = {arXiv},
       eprint = {1708.05519},
 primaryClass = {astro-ph.GA},
       adsurl = {https://ui.adsabs.harvard.edu/abs/2018MNRAS.474.3615M}
}

@ARTICLE{2001MNRAS.328..185S,
       author = {{Schulz}, Steven and {Struck}, Curtis},
        title = "{Multi stage three-dimensional sweeping and annealing of disc galaxies in clusters}",
      journal = {\mnras},
         year = 2001,
        month = nov,
       volume = {328},
       number = {1},
        pages = {185-202},
          doi = {10.1046/j.1365-8711.2001.04847.x},
archivePrefix = {arXiv},
       eprint = {astro-ph/0107570},
 primaryClass = {astro-ph},
       adsurl = {https://ui.adsabs.harvard.edu/abs/2001MNRAS.328..185S}
}

@ARTICLE{2009ApJ...694..789T,
       author = {{Tonnesen}, Stephanie and {Bryan}, Greg L.},
        title = "{Gas Stripping in Simulated Galaxies with a Multiphase Interstellar Medium}",
      journal = {\apj},
         year = 2009,
        month = apr,
       volume = {694},
       number = {2},
        pages = {789-804},
          doi = {10.1088/0004-637X/694/2/789},
archivePrefix = {arXiv},
       eprint = {0901.2115},
 primaryClass = {astro-ph.GA},
       adsurl = {https://ui.adsabs.harvard.edu/abs/2009ApJ...694..789T}
}

@ARTICLE{2018MNRAS.476.3781R,
       author = {{Ramos-Mart{\'\i}nez}, Mariana and {G{\'o}mez}, Gilberto C. and {P{\'e}rez-Villegas}, {\'A}ngeles},
        title = "{MHD simulations of ram pressure stripping of a disc galaxy}",
      journal = {\mnras},
         year = 2018,
        month = may,
       volume = {476},
       number = {3},
        pages = {3781-3792},
          doi = {10.1093/mnras/sty393},
archivePrefix = {arXiv},
       eprint = {1711.01252},
 primaryClass = {astro-ph.GA},
       adsurl = {https://ui.adsabs.harvard.edu/abs/2018MNRAS.476.3781R}
}

@ARTICLE{2014MNRAS.443.1151N,
       author = {{Norris}, Mark A. and {Kannappan}, Sheila J. and {Forbes}, Duncan A. and {Romanowsky}, Aaron J. and {Brodie}, Jean P. and {Faifer}, Favio Ra{\'u}l and {Huxor}, Avon and {Maraston}, Claudia and {Moffett}, Amanda J. and {Penny}, Samantha J. and {Pota}, Vincenzo and {Smith-Castelli}, Anal{\'\i}a and {Strader}, Jay and {Bradley}, David and {Eckert}, Kathleen D. and {Fohring}, Dora and {McBride}, JoEllen and {Stark}, David V. and {Vaduvescu}, Ovidiu},
        title = "{The AIMSS Project - I. Bridging the star cluster-galaxy divide$^{★}${\textdagger}{\textdaggerdbl}{\textsection}{\textparagraph}}",
      journal = {\mnras},
         year = 2014,
        month = sep,
       volume = {443},
       number = {2},
        pages = {1151-1172},
          doi = {10.1093/mnras/stu1186},
archivePrefix = {arXiv},
       eprint = {1406.6065},
 primaryClass = {astro-ph.GA},
       adsurl = {https://ui.adsabs.harvard.edu/abs/2014MNRAS.443.1151N}
}

@ARTICLE{2015MNRAS.451.3615N,
       author = {{Norris}, Mark A. and {Escudero}, Carlos G. and {Faifer}, Favio R. and {Kannappan}, Sheila J. and {Forte}, Juan Carlos and {van den Bosch}, Remco C.~E.},
        title = "{An extended star formation history in an ultra-compact dwarf}",
      journal = {\mnras},
         year = 2015,
        month = aug,
       volume = {451},
       number = {4},
        pages = {3615-3626},
          doi = {10.1093/mnras/stv1221},
archivePrefix = {arXiv},
       eprint = {1506.00004},
 primaryClass = {astro-ph.GA},
       adsurl = {https://ui.adsabs.harvard.edu/abs/2015MNRAS.451.3615N}
}

@ARTICLE{2016MNRAS.456..617J,
       author = {{Janz}, Joachim and {Norris}, Mark A. and {Forbes}, Duncan A. and {Huxor}, Avon and {Romanowsky}, Aaron J. and {Frank}, Matthias J. and {Escudero}, Carlos G. and {Faifer}, Favio R. and {Forte}, Juan Carlos and {Kannappan}, Sheila J. and {Maraston}, Claudia and {Brodie}, Jean P. and {Strader}, Jay and {Thompson}, Bradley R.},
        title = "{The AIMSS Project - III. The stellar populations of compact stellar systems}",
      journal = {\mnras},
         year = 2016,
        month = feb,
       volume = {456},
       number = {1},
        pages = {617-632},
          doi = {10.1093/mnras/stv2636},
archivePrefix = {arXiv},
       eprint = {1511.03264},
 primaryClass = {astro-ph.GA},
       adsurl = {https://ui.adsabs.harvard.edu/abs/2016MNRAS.456..617J}
}

@ARTICLE{1998RvMP...70....1B,
       author = {{Balbus}, Steven A. and {Hawley}, John F.},
        title = "{Instability, turbulence, and enhanced transport in accretion disks}",
      journal = {Reviews of Modern Physics},
         year = 1998,
        month = jan,
       volume = {70},
       number = {1},
        pages = {1-53},
          doi = {10.1103/RevModPhys.70.1},
       adsurl = {https://ui.adsabs.harvard.edu/abs/1998RvMP...70....1B}
}

@ARTICLE{2009MNRAS.395.2183Y,
       author = {{Yuan}, Feng and {Lin}, Jun and {Wu}, Kinwah and {Ho}, Luis C.},
        title = "{A magnetohydrodynamical model for the formation of episodic jets}",
      journal = {\mnras},
         year = 2009,
        month = jun,
       volume = {395},
       number = {4},
        pages = {2183-2188},
          doi = {10.1111/j.1365-2966.2009.14673.x},
archivePrefix = {arXiv},
       eprint = {0811.2893},
 primaryClass = {astro-ph},
       adsurl = {https://ui.adsabs.harvard.edu/abs/2009MNRAS.395.2183Y}
}

@ARTICLE{1991ApJ...376..214B,
       author = {{Balbus}, Steven A. and {Hawley}, John F.},
        title = "{A Powerful Local Shear Instability in Weakly Magnetized Disks. I. Linear Analysis}",
      journal = {\apj},
         year = 1991,
        month = jul,
       volume = {376},
        pages = {214},
          doi = {10.1086/170270},
       adsurl = {https://ui.adsabs.harvard.edu/abs/1991ApJ...376..214B}
}

@ARTICLE{2025ApJ...991...89C,
       author = {{Cho}, Hyerin and {Narayan}, Ramesh},
        title = "{Variability in Black Hole Accretion: Dependence on Rotational and Magnetic Energy Balance}",
      journal = {\apj},
         year = 2025,
        month = sep,
       volume = {991},
       number = {1},
          eid = {89},
        pages = {89},
          doi = {10.3847/1538-4357/adf8d3},
archivePrefix = {arXiv},
       eprint = {2507.13441},
 primaryClass = {astro-ph.HE},
       adsurl = {https://ui.adsabs.harvard.edu/abs/2025ApJ...991...89C}
}

@ARTICLE{2022ApJ...941...30C,
       author = {{Chatterjee}, K. and {Narayan}, R.},
        title = "{Flux Eruption Events Drive Angular Momentum Transport in Magnetically Arrested Accretion Flows}",
      journal = {\apj},
         year = 2022,
        month = dec,
       volume = {941},
       number = {1},
          eid = {30},
        pages = {30},
          doi = {10.3847/1538-4357/ac9d97},
archivePrefix = {arXiv},
       eprint = {2210.08045},
 primaryClass = {astro-ph.HE},
       adsurl = {https://ui.adsabs.harvard.edu/abs/2022ApJ...941...30C}
}

@ARTICLE{2011MNRAS.418L..79T,
       author = {{Tchekhovskoy}, Alexander and {Narayan}, Ramesh and {McKinney}, Jonathan C.},
        title = "{Efficient generation of jets from magnetically arrested accretion on a rapidly spinning black hole}",
      journal = {\mnras},
         year = 2011,
        month = nov,
       volume = {418},
       number = {1},
        pages = {L79-L83},
          doi = {10.1111/j.1745-3933.2011.01147.x},
archivePrefix = {arXiv},
       eprint = {1108.0412},
 primaryClass = {astro-ph.HE},
       adsurl = {https://ui.adsabs.harvard.edu/abs/2011MNRAS.418L..79T}
}

@ARTICLE{2008MNRAS.383..458C,
       author = {{Cuadra}, Jorge and {Nayakshin}, Sergei and {Martins}, Fabrice},
        title = "{Variable accretion and emission from the stellar winds in the Galactic Centre}",
      journal = {\mnras},
         year = 2008,
        month = jan,
       volume = {383},
       number = {2},
        pages = {458-466},
          doi = {10.1111/j.1365-2966.2007.12573.x},
archivePrefix = {arXiv},
       eprint = {0705.0769},
 primaryClass = {astro-ph},
       adsurl = {https://ui.adsabs.harvard.edu/abs/2008MNRAS.383..458C}
}

@ARTICLE{2019Natur.570...83M,
       author = {{Murchikova}, Elena M. and {Phinney}, E. Sterl and {Pancoast}, Anna and {Blandford}, Roger D.},
        title = "{A cool accretion disk around the Galactic Centre black hole}",
      journal = {\nat},
         year = 2019,
        month = jun,
       volume = {570},
       number = {7759},
        pages = {83-86},
          doi = {10.1038/s41586-019-1242-z},
archivePrefix = {arXiv},
       eprint = {1906.08289},
 primaryClass = {astro-ph.GA},
       adsurl = {https://ui.adsabs.harvard.edu/abs/2019Natur.570...83M}
}

@ARTICLE{2009ApJ...694..556B,
       author = {{Blakeslee}, John P. and {Jord{\'a}n}, Andr{\'e}s and {Mei}, Simona and {C{\^o}t{\'e}}, Patrick and {Ferrarese}, Laura and {Infante}, Leopoldo and {Peng}, Eric W. and {Tonry}, John L. and {West}, Michael J.},
        title = "{The ACS Fornax Cluster Survey. V. Measurement and Recalibration of Surface Brightness Fluctuations and a Precise Value of the Fornax-Virgo Relative Distance}",
      journal = {\apj},
         year = 2009,
        month = mar,
       volume = {694},
       number = {1},
        pages = {556-572},
          doi = {10.1088/0004-637X/694/1/556},
archivePrefix = {arXiv},
       eprint = {0901.1138},
 primaryClass = {astro-ph.CO},
       adsurl = {https://ui.adsabs.harvard.edu/abs/2009ApJ...694..556B}
}

@ARTICLE{2020ApJ...888L...2C,
       author = {{Calder{\'o}n}, Diego and {Cuadra}, Jorge and {Schartmann}, Marc and {Burkert}, Andreas and {Russell}, Christopher M.~P.},
        title = "{Stellar Winds Pump the Heart of the Milky Way}",
      journal = {\apjl},
         year = 2020,
        month = jan,
       volume = {888},
       number = {1},
          eid = {L2},
        pages = {L2},
          doi = {10.3847/2041-8213/ab5e81},
archivePrefix = {arXiv},
       eprint = {1910.06976},
 primaryClass = {astro-ph.GA},
       adsurl = {https://ui.adsabs.harvard.edu/abs/2020ApJ...888L...2C}
}

@ARTICLE{2010MNRAS.403.1413K,
       author = {{Karakas}, A.~I.},
        title = "{Updated stellar yields from asymptotic giant branch models}",
      journal = {\mnras},
         year = 2010,
        month = apr,
       volume = {403},
       number = {3},
        pages = {1413-1425},
          doi = {10.1111/j.1365-2966.2009.16198.x},
archivePrefix = {arXiv},
       eprint = {0912.2142},
 primaryClass = {astro-ph.SR},
       adsurl = {https://ui.adsabs.harvard.edu/abs/2010MNRAS.403.1413K}
}

@ARTICLE{2015ApJS..219...40C,
       author = {{Cristallo}, S. and {Straniero}, O. and {Piersanti}, L. and {Gobrecht}, D.},
        title = "{Evolution, Nucleosynthesis, and Yields of AGB Stars at Different Metallicities. III. Intermediate-mass Models, Revised Low-mass Models, and the ph-FRUITY Interface}",
      journal = {\apjs},
         year = 2015,
        month = aug,
       volume = {219},
       number = {2},
          eid = {40},
        pages = {40},
          doi = {10.1088/0067-0049/219/2/40},
archivePrefix = {arXiv},
       eprint = {1507.07338},
 primaryClass = {astro-ph.SR},
       adsurl = {https://ui.adsabs.harvard.edu/abs/2015ApJS..219...40C}
}

@ARTICLE{2018MNRAS.475.2282V,
       author = {{Ventura}, P. and {Karakas}, A. and {Dell'Agli}, F. and {Garc{\'\i}a-Hern{\'a}ndez}, D.~A. and {Guzman-Ramirez}, L.},
        title = "{Gas and dust from solar metallicity AGB stars}",
      journal = {\mnras},
         year = 2018,
        month = apr,
       volume = {475},
       number = {2},
        pages = {2282-2305},
          doi = {10.1093/mnras/stx3338},
archivePrefix = {arXiv},
       eprint = {1712.08582},
 primaryClass = {astro-ph.SR},
       adsurl = {https://ui.adsabs.harvard.edu/abs/2018MNRAS.475.2282V}
}
\bibliographystyle{aasjournalv7}



\end{document}